\documentclass[a4paper,fleqn,usenatbib,useAMS]{mnras}

\usepackage{graphicx}	
\usepackage{amsmath}	
\usepackage{amssymb}	
\usepackage{multicol}        
\usepackage{bm}		
\usepackage{pdflscape}	
\usepackage{graphicx}
\usepackage{xcolor}
\usepackage{subfig}
\usepackage{caption}

\definecolor{purple}{rgb}{0.4,0.0,1.} 
\definecolor{lightgreen}{rgb}{0.67,0.87,0.0}

\usepackage[T1]{fontenc}
\usepackage{ae,aecompl}

\usepackage{newtxtext,newtxmath}

\title[NGFS:~VII:. A MUSE view of the NSCs in dwarf galaxies]{The Next Generation Fornax Survey (NGFS):~VII.~A MUSE view of the nuclear star clusters in Fornax dwarf galaxies}
\author[E. J. Johnston \textit{et al}]{Evelyn J. Johnston$^{1}$\thanks{Contact e-mail: \href{mailto:ejohnston@astro.puc.cl}{ejohnston@astro.puc.cl} }, Thomas H.~Puzia$^{1}$, Giuseppe D'Ago$^{1}$, Paul Eigenthaler$^{1}$,  
\newauthor  Gaspar Galaz$^{1}$, Boris H\" au\ss ler$^{2}$, Marcelo D. Mora$^{1}$,  Yasna Ordenes-Brice\~no$^{1}$, 
\newauthor  Yu Rong$^{1}$, Chelsea Spengler$^{1}$,  Fr\'ed\'eric Vogt$^{2}$, Patrick C{\^o}t{\'e}$^{3}$, Eva K.\ Grebel$^{4}$, 
\newauthor   Michael Hilker$^{5}$,Steffen Mieske$^{2}$, Bryan Miller$^{6}$, Ruben S\'anchez-Janssen$^{7}$,  
\newauthor  Matthew A. Taylor$^{3,8}$, \&  Hong-Xin Zhang$^{9,10}$ \\ 
$^{1}$Institute of Astrophysics, Pontificia Universidad Cat\'olica de Chile, Av.~Vicu\~na Mackenna 4860, 7820436 Macul, Santiago, Chile\\
$^{2}$European Southern Observatory, Alonso de C\'ordova 3107, Vitacura, Santiago \\
$^{3}$ National Research Council of Canada, Herzberg Astronomy and Astrophysics Research Centre, 5701 West Saanich Road, Victoria, BC V9E 2E7, Canada\\
$^{4}$Astronomisches Rechen-Institut, Zentrum f\"ur Astronomie der Universit\"at Heidelberg, M\"onchhofstra{\ss}e 12-14, D-69120 Heidelberg, Germany\\
$^{5}$European Southern Observatory, Karl-Schwarzchild-Str. 2, D-85748 Garching, Germany\\
$^{6}$Gemini Observatory, South Operations Center, Casilla 603, La Serena, Chile \\
$^{7}$STFC UK Astronomy Technology Centre, Royal Observatory, Blackford Hill, Edinburgh, EH9 3HJ, UK \\
$^{8}$ Gemini Observatory, Northern Operations Center, 670 North A'ohoku Place, Hilo, HI 96720-1906, USA\\
$^{9}$ CAS Key Laboratory for Research in Galaxies and Cosmology, Department of Astronomy, University of Science and Technology of China, Hefei, Anhui 230026, China\\
$^{10}$ School of Astronomy and Space Science, University of Science and Technology of China, Hefei 230026, China\\
}

\pubyear{2020?}
\begin{document}
\label{firstpage}
\pagerange{\pageref{firstpage}--\pageref{lastpage}}
\maketitle

\begin{abstract}
Clues to the formation and evolution of Nuclear Star Clusters (NSCs) lie in their stellar populations. However, these structures are often very faint compared to their host galaxy, and spectroscopic analysis of NSCs is hampered by contamination of light from the rest of the system. With the introduction of wide-field IFU spectrographs, new techniques have been developed to model the light from different components within galaxies, making it possible to cleanly extract the spectra of the NSCs and study their properties with minimal contamination from the light of the rest of the galaxy. This work presents the analysis of the NSCs in a sample of 12 dwarf galaxies in the Fornax Cluster observed with MUSE. Analysis of the stellar populations and star-formation histories reveal that \textit{all the NSCs  show evidence of multiple episodes of star formation, indicating that they have built up their mass further since their initial formation}. The NSCs were found to have systematically lower metallicities than their host galaxies, which is consistent with a scenario for mass-assembly through mergers with infalling globular clusters, while the presence of younger stellar populations and gas emission in the core of two galaxies is indicative of in-situ star formation. We conclude that the NSCs in these dwarf galaxies likely originated as globular clusters that migrated to the core of the galaxy which have built up their mass mainly through mergers with other infalling clusters, with gas-inflow leading to in-situ star formation playing a secondary role.
\end{abstract}


\section{Introduction}\label{sec:intro}

Nuclear Star Clusters (NSCs) are compact stellar systems lying at the centres of galaxies with similar sizes to globular clusters (GCs) but a broader range of masses, typically from $10^5-10^8M_{\odot}$ \citep{Walcher_2006, Georgiev_2016, Spengler_2017, Neumayer_2020}. They are a common characteristic of galaxies of all morphologies and with masses $>10^6M_{\odot}$, and their occurrence is linked to the mass of the host galaxy. Deep imaging surveys of the Virgo Cluster have shown that the nucleation fraction increases with the galaxy mass up to $M_*\sim10^{9.5}M_{\odot}$, and then decreases towards higher masses \citep{SanchezJanssen_2019}.  \citet{Cote_2006} explained the lack of NSCs in the highest-mass galaxies with the proposal that  NSCs are the low-mass counterparts of supermassive black holes (SMBHs) in these galaxies, and thus that the growth of their stellar-mass black holes leads to their disruption as the SMBH sphere of influence reaches the NSC size. Furthermore, the properties of the NSCs appear to be linked to those of the host galaxies through various scaling relations, such as their masses, stellar populations and colours \citep[e.g.][]{Walcher_2005, Turner_2012, Scott_2013, Georgiev_2014, Georgiev_2016}. For example, starting at lowest galaxy masses the mass fraction of the NSCs has been found to decrease with increasing galaxy mass up to $M_*\approx10^{9.5}M_{\odot}$, before increasing again at higher masses \citep{Ordenes_2018b, SanchezJanssen_2019}. 

Two main scenarios have been proposed to explain the formation of NSCs. The first process is the \textit{in situ} scenario, in which the NSC is formed in the central few parsecs of the galaxy through star formation fuelled by infalling gas \citep{Milosavljevic_2004, Bekki_2006, Bekki_2007}. In this scenario, the star formation timescale and occurrences of later star formation are dependent on the gas reservoirs within the galaxy. The second scenario is the \textit{migration} scenario, in which a GC forms  within the galaxy and migrates to the centre of the galaxy via dynamical friction \citep{Tremaine_1975}. NSCs created in this way can build up their mass at later times through mergers with other infalling GCs \citep{Andersen_2008}. 

A hybrid scenario has also been suggested- the so-called \textit{wet-migration} scenario \citep{Guillard_2016}. In this case, a young, massive star cluster (YMSC) forms with its own gas reservoir somewhere in the disc of the galaxy. As the YMSC migrates to the centre of the galaxy through dynamical friction and interactions with other structures, the gas reservoir feeds the ongoing star formation, leading to the YMSC growing in mass as it migrates and settles at the core of the galaxy.

Clues to the formation and evolution of NSCs and their host galaxies lie in their stellar populations. \citet{Paudel_2011} used long-slit spectra to study the stellar populations of NSCs in dwarf ellipticals (dEs) in the Virgo Cluster, finding that in general the NSCs were younger and more metal-rich than their host galaxies. They interpreted these trends to indicate that the young, metal-enriched NSCs formed at later times in the lifetime of the host galaxy, or that the NSCs formed through an extended period of star formation fuelled by gas inflow. Another photometric survey of dEs in the Virgo Cluster by \citet{Spengler_2017}, however, found that while the NSCs were bluer in optical colours than their host galaxies, the NSC ages showed a wide distribution with no clear age or metallicity differences when compared to the host galaxies. \citet{Lotz_2004} compared the colours of NSCs with globular clusters within the same galaxy, finding them to be similar or slightly redder. They also identified two very blue NSCs, which they attributed to recent star formation. \citet{Fahrion_2019} on the other hand extracted the spectrum of the NSC of the elliptical galaxy FCC~47, and found it to contain metal-rich but very old stars. Additionally, many studies have found evidence of multiple stellar populations within NSCs \citep[e.g.][]{Walcher_2006, Rossa_2006, Seth_2006, Seth_2010, Koleva_2011, Lyubenova_2013}, where generally the younger stars appear to be more concentrated within the core of the NSC while the older stars have a more extended distribution. However, a study by \citet{Carson_2015} found two galaxies with extended circumnuclear star formation within the NSC.

It is unlikely that one single process dominates the formation of NSCs. MUSE observations of the NSC in the Sagittarius dwarf galaxy by \citet{AlfaroCuello_2019}, in which the individual stars could be resolved and studied spectroscopically, revealed both young, metal-rich and old, metal-poor stellar populations, and proposed that the latter population was created through inward migration of GCs while the younger population  were formed through in-situ star formation fuelled by metal-rich gas. \citet{Fahrion_2020} also found evidence of lower metallicities in the NSCs of two dwarf galaxies, which they proposed was consistent with the NSC forming via in-spiraling GCs. \citet{Walcher_2005}, \citet{Misgeld_2011} and \citet{Norris_2014} further reflected that NSCs have the highest stellar densities in the universe, and that while mergers of GCs may play an important role in building up their mass, in-situ star formation is required to increase the central stellar density to the observed levels \citep{Neumayer_2017}. It is also widely thought that the mass of the host galaxy plays an important role in the process by which the NSC builds up its mass. \citet{Turner_2012} and \citet{den_brok_2014} found evidence that mergers with migrating clusters are responsible for the growth of the NSC in low-mass galaxies, with in-situ star formation playing an important secondary role. They also found that in higher-mass galaxies, gas accretion from mergers becomes increasingly important and contributes to the in-situ star formation. \citet{Antonini_2015} elaborated on this theory with simulations that showed that the in-situ star-formation becomes increasingly dominant above a host galaxy mass of $3\times10^{11}M_{\odot}$, while \citet{Lyubenova_2019} found that they were unable to rule out completely the pure GC merging scenario in galaxies with masses $>10^{9.7}M_{\odot}$. Furthermore, \citet{SanchezJanssen_2019} concluded that the formation of NSCs in dwarf galaxies is largely stochastic, and noted that there appears to be a close connection between NSCs and GCs in terms of mass and occupation distributions at lower host galaxy masses.~An earlier study by \citet{Lotz_2001} of dwarf galaxies in the Fornax and Virgo clusters found a deficit of bright GCs in the inner regions of many galaxies, which they interpreted as evidence of orbital decay of the missing GCs that have spiralled into the core to form the NSC. Consequently, it is becoming increasingly important to not only identify the dominant processes that built up the mass of the NSC, but instead to consider the relative importance of the dissipational and dissipationless processes as a function of mass and environment.

Spectroscopic studies of the stellar populations of NSCs are challenging due to the faintness of the NSCs relative to their host galaxies, and the resulting difficulties of extracting their spectra while minimizing the contamination from the light of the rest of the galaxy.  However, with the wide field-of-view and high spatial resolution offered by modern Integral Field Unit (IFU) spectrographs, it is now possible to model the 2-dimensional light profile of a NSC and its host galaxy as a function of wavelength in order to cleanly extract their integrated spectra. One such code designed to carry out this modelling and extraction process is \textsc{buddi} \citep[Bulge--Disc Decomposition of IFU data,][]{Johnston_2017}. Whilst originally designed to model bulges and discs of galaxies, \textsc{buddi} can be used to extract the spectra of any structures that can be modelled with a smooth, symmetric light profile, such as bars, globular clusters and extended stellar haloes \citep{Johnston_2018}.
In this paper, we use this novel approach to create wavelength-dependent models of a sample of nucleated dwarf galaxies in the Fornax Cluster. The complexity of the models was built up until the best fit to the galaxy was obtained, thus ensuring minimal contamination of the spectrum of the NSC from the light of the host galaxy. These spectra were then used to study the stellar populations and star-formation histories of the NSCs in order to determine how they have formed and built up their mass.

This paper is laid out as follows: Section~\ref{sec:DR} describes the sample selection, observations, and data reduction;  Section~\ref{sec:BUDDI} outlines the spectroscopic decomposition technique used to cleanly extract the spectrum of the NSC; Section~\ref{sec:analysis1} contains the analysis of the physical parameters for the NSCs and their host galaxies; Sections~\ref{sec:analysis2} and \ref{sec:analysis3} present the stellar population analysis through line-strength analysis and full spectral fitting, respectively; and finally Section~\ref{sec:discussion} discusses the results and presents the conclusions. Throughout this paper we assume 
a distance modulus of $31.51 \pm 0.03$~mag for Fornax, which corresponds to a distance of 20~Mpc \citep{Blakeslee_2009}.


\section{Data and Sample}\label{sec:DR}
\subsection{Sample Selection}\label{sec:sample}
The initial sample selection for this study comprised of all the nucleated dwarf galaxies in the core Fornax Cluster according to the Fornax Dwarf Galaxy Catalog \citep[$M<10^9M_\odot$;][]{Venhola_2018} that have been observed with the Multi-Unit Spectroscopic Explorer \citep[MUSE;][]{Bacon_2010} at the Very Large Telescope (VLT), and which are publicly available in the ESO Archive Facility. Using these criteria, a total of 11 galaxies were identified in the Archive, and a twelfth galaxy, FCC~207, was added to the sample as it was classified as a nucleated dwarf by \citet{DeRijcke_2003}. These galaxies were observed between October 2014 and January 2019, using the wide-field mode (WFM) in both nominal and extended modes, with and without AO. Full details of the observations are given in Table~\ref{tab:data}.

\begin{table*}
	\centering
	\caption{Observations }
	\label{tab:data}
	\begin{tabular}{lrrrrrr} 
		\hline
		Galaxy &   & Program  & PI & Exposure Time  & Mode & Date \\
		\hline
FCC~182 & & 296.B-5054 &  Sarzi &  $5\times720$s & WFM-NOAO-E & Jan. 2017\\
                & &                     &  Sarzi &  $3\times600$s & WFM-NOAO-E  & Jul. 2017\\
FCC~188 & & 096.B-0399 &  Napolitano &  $8\times1300$s  & WFM-NOAO-N & Oct., Dec. 2015; Jan. 2016\\
FCC~202$^*$& & 094.B-0895  &  Lisker &  $16\times1300$s & WFM-NOAO-N & Dec. 2014, Jan. 2015 \\
FCC~207 & & 094.B-0576 & Emsellem  &  $5\times600$s & WFM-NOAO-N &  Oct. 2014\\
                & & 096.B-0063  & Emsellem &  $8\times600$s & WFM-NOAO-N  & Nov. 2015\\
                & & 097.B-0761  & Emsellem &  $4\times600$s & WFM-NOAO-N &  Sept. 2016\\
                & & 098.B-0239  & Emsellem &  $4\times600$s & WFM-NOAO-N  & Jan. 2017\\
FCC~211 & & 096.B-0399 &  Napolitano &  $8\times1300$s & WFM-NOAO-N &  Oct., Dec. 2015; Jan. 2016\\
FCC~215 & & 096.B-0399 &  Napolitano &  $8\times1300$s & WFM-NOAO-N  &  Oct., Dec. 2015; Jan. 2016\\
FCC~222 & & 096.B-0399 &  Napolitano &  $8\times1300$s & WFM-NOAO-N  &  Oct., Dec. 2015; Jan. 2016\\
FCC~223 & & 096.B-0399 &  Napolitano &  $8\times1300$s & WFM-NOAO-N  &  Nov., Dec. 2015; Jan. 2016\\
FCC~227 & & 096.B-0399 & Napolitano &   $8\times1300$s & WFM-NOAO-N &  Nov., Dec. 2015; Jan. 2016\\
FCC~245 & &101.C-0329 & Vogt &   $2\times675$s & WFM-NOAO-N & Jul., Oct. 2018 \\
FCC~306 & &296.B-5054 &  Sarzi &  $4\times900$s & WFM-NOAO-E & Oct., Dec 2016\\
FCCB~1241$^{**}$ & &102.B-0455 &  Johnston &  $3\times720$s & WFM-AO-N  & Jan. 2019\\
		\hline
\multicolumn{7}{p{5.0in}}{$^*$ Observations of FCC~202 were split up over two pointings to cover the whole galaxy. \newline
$^{**}$ FCCB~1241 was originally classified as a background galaxy in the Fornax Cluster Catalog \citep{Ferguson_1989}, but has since been spectroscopically confirmed as a member of the Fornax Cluster by \citet{Deady_2002}. The FCCB prefix will be used throughout this paper as no FCC prefix has been assigned to this galaxy.}
	\end{tabular}
\end{table*}

\begin{figure}
\begin{center}
\includegraphics[angle=0,width=\linewidth]{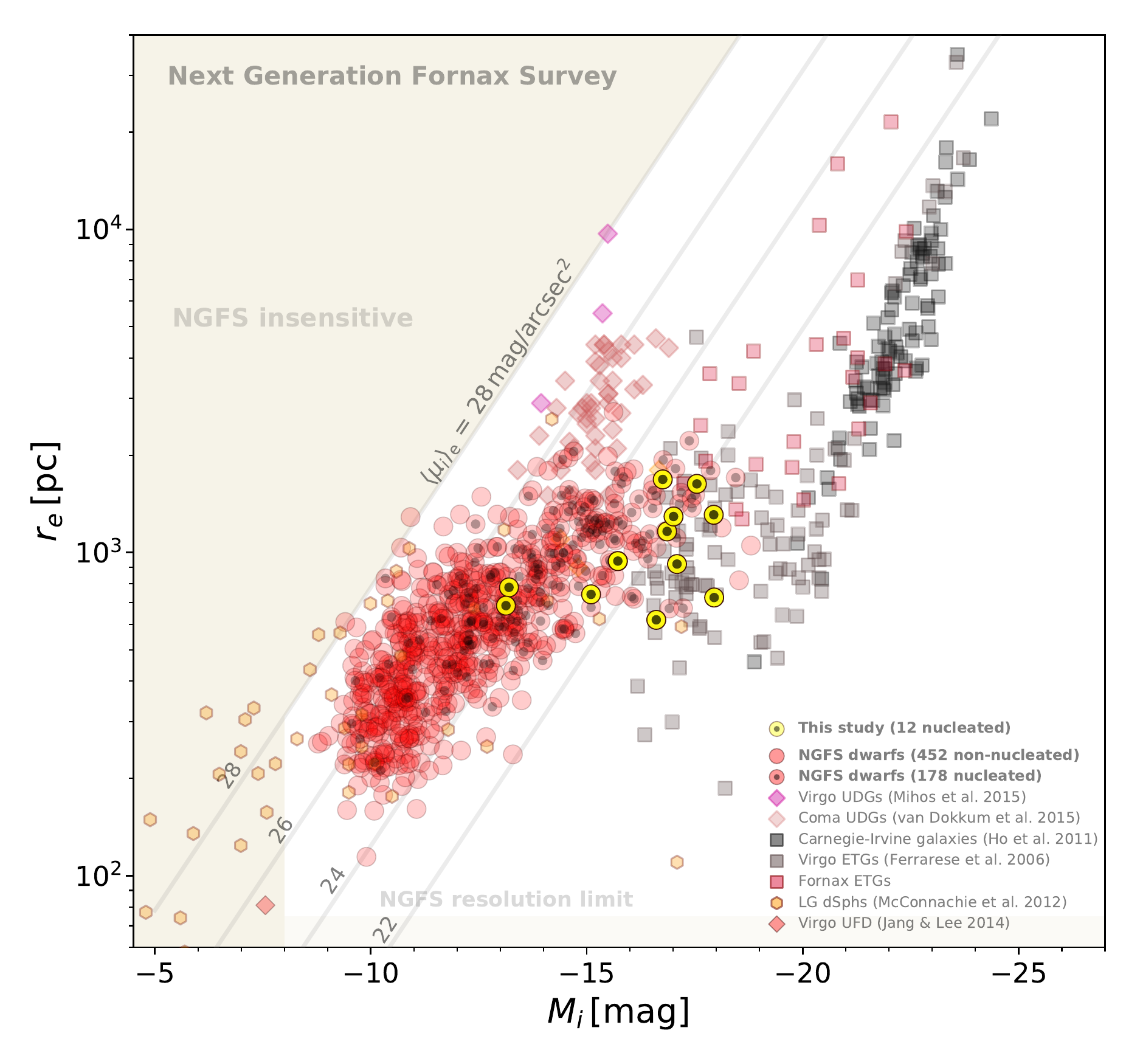}
\caption{The magnitude-size relation of \citet{Ordenes_2018a}, showing the Fornax dwarf galaxies detected by the NGFS within $R_{\rm vir}/2$ as red circles and with the sample of galaxies used in this work highlighted as yellow circles. 
 \label{fig:mass_size}}
\end{center}
\end{figure}

The magnitude-size relation for galaxies in the Fornax cluster from \citet{Ordenes_2018a} is given in Fig.~\ref{fig:mass_size}, and reveals that the majority of these galaxies lie towards the bright end of the dwarf-galaxy sequence, with a few fainter galaxies included in the sample. All but one galaxy lie within the core of the Fornax Cluster, defined in this paper as the $R_{\rm vir}/4$ radius on the sky, centred on NGC\,1399 ($R\approx350$~kpc), and which corresponds to the central tile of the Next Generation Fornax Survey \citep[NGFS Tile~1,][]{Munoz_2015}. The remaining galaxy, FCC~306, lies within the $R_{\rm vir}/2$, within the NGFS Tile~6. Additionally, 10 of the galaxies show no evidence of gas, while FCC~207 shows evidence of gas emission at its core and FCC~306 contains gas distributed throughout the galaxy.

\subsection{Spectroscopic data}\label{sec:MUSE_data}

MUSE is an optical integral-field spectrograph with a field of view of $1\arcmin\!\times\!1\arcmin$, a spatial resolution of 0.2\arcsec/pixel, and a spectral resolving power ranging from \textit{R}$\simeq$1770 at 4800\,\AA\ to \textit{R}$\simeq$3590 at 9300\,\AA. The data were reduced using the ESO MUSE pipeline \citep[v2.6,][]{Weilbacher_2012} in the ESO Recipe Execution Tool (EsoRex) environment \citep{ESOREX}.~Master bias, flat field and wavelength calibrations were created for each night of observations from the associated raw calibrations, and were applied to the raw science and standard-star observations as part of the pre-processing steps. A flux calibration solution was obtained for each night using exposures of standard stars from the same night, and the sky continuum was measured for each pair of science exposures using a separate sky exposure where appropriate, and using the edges of the field-of-view of the science exposure in cases where a  sky exposure was not taken. The post-processing step of the data reduction applied the flux calibration and sky subtraction to the science exposures, and the data for each galaxy were  combined to produce the final datacube. As a final step, the residual sky contamination was cleaned using the Zurich Atmosphere Purge code \citep[ZAP, ][]{Soto_2016}. 

\begin{figure}
\begin{center}
\includegraphics[angle=0,width=0.30\linewidth]{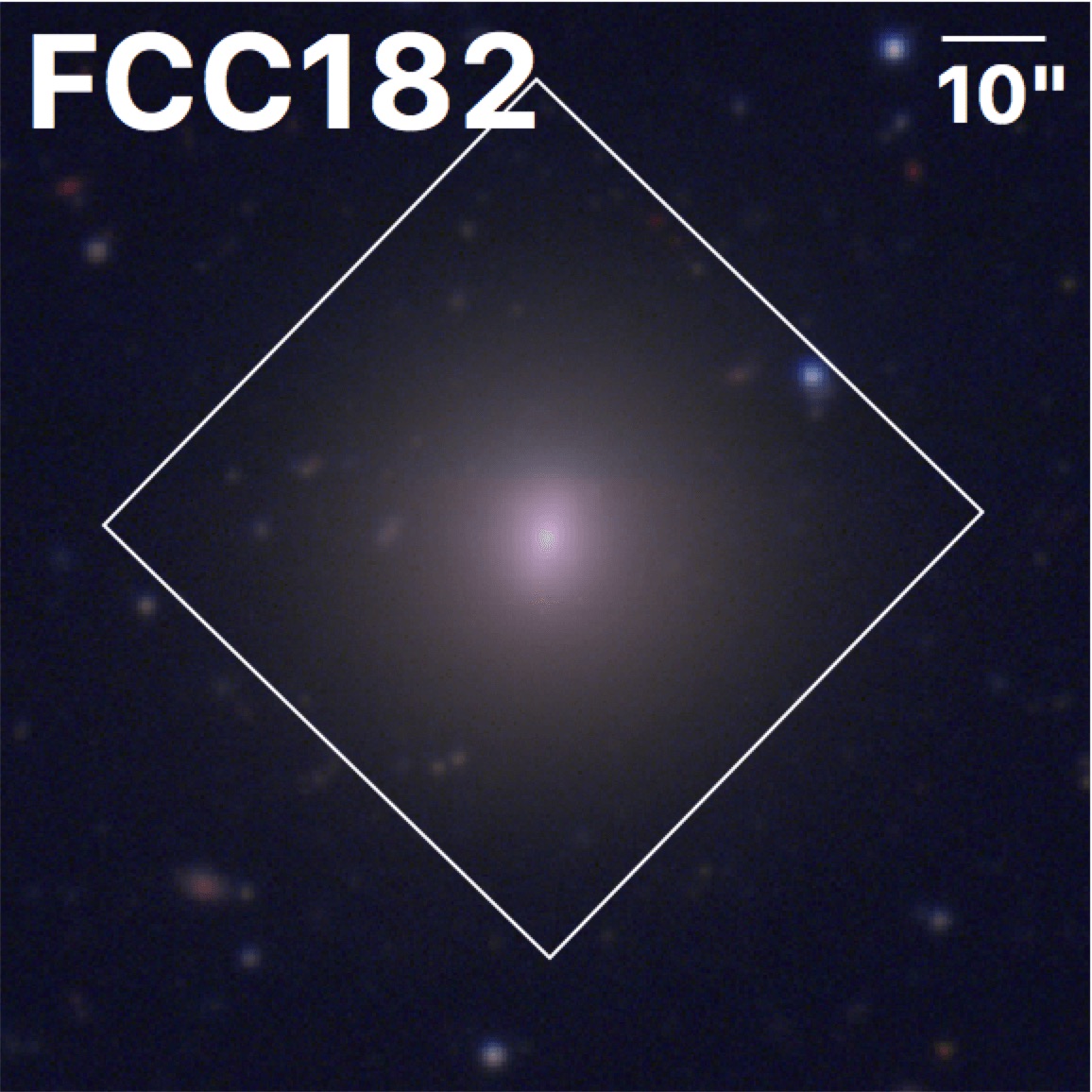}
\includegraphics[angle=0,width=0.30\linewidth]{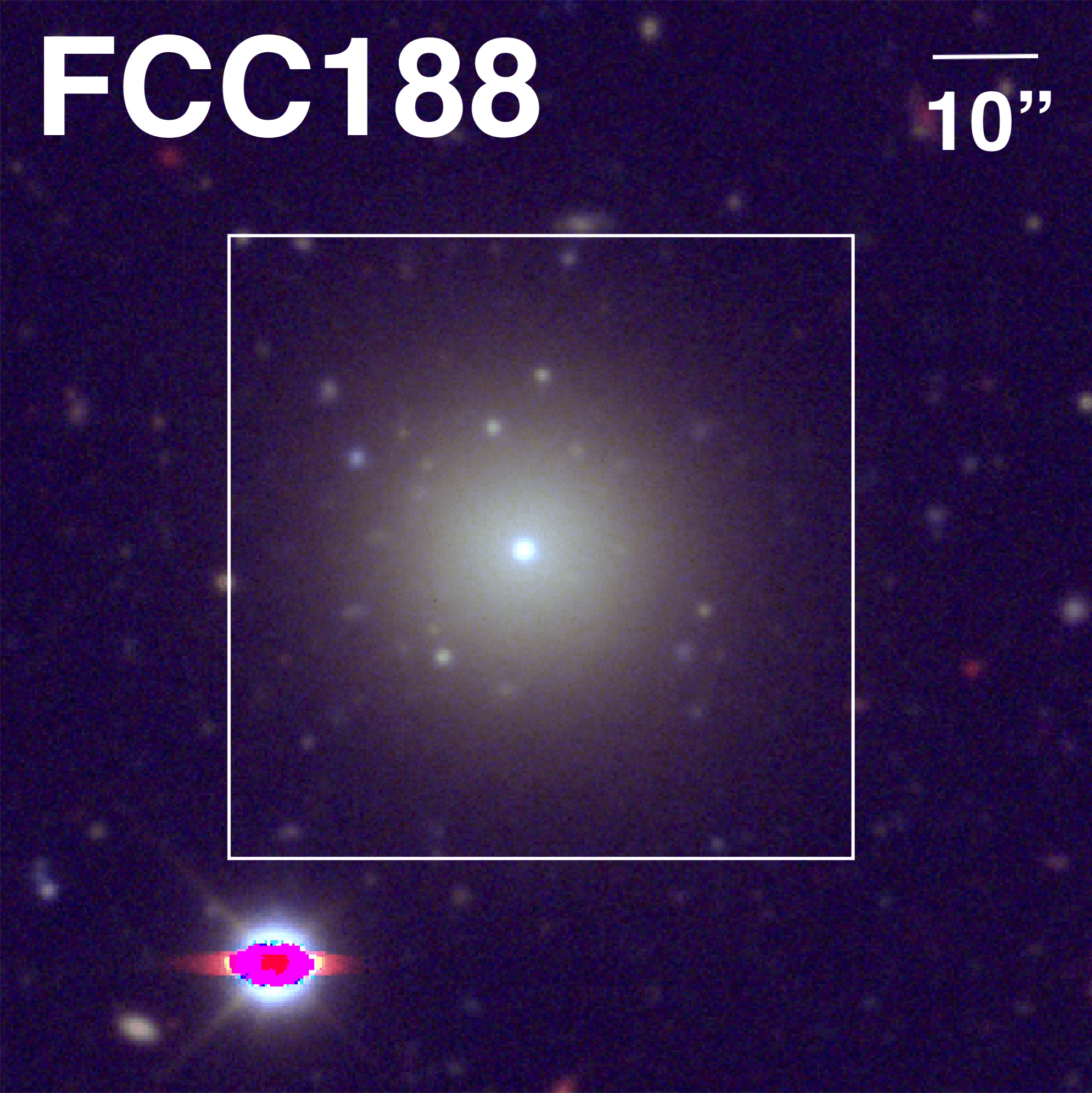}
\includegraphics[angle=0,width=0.30\linewidth]{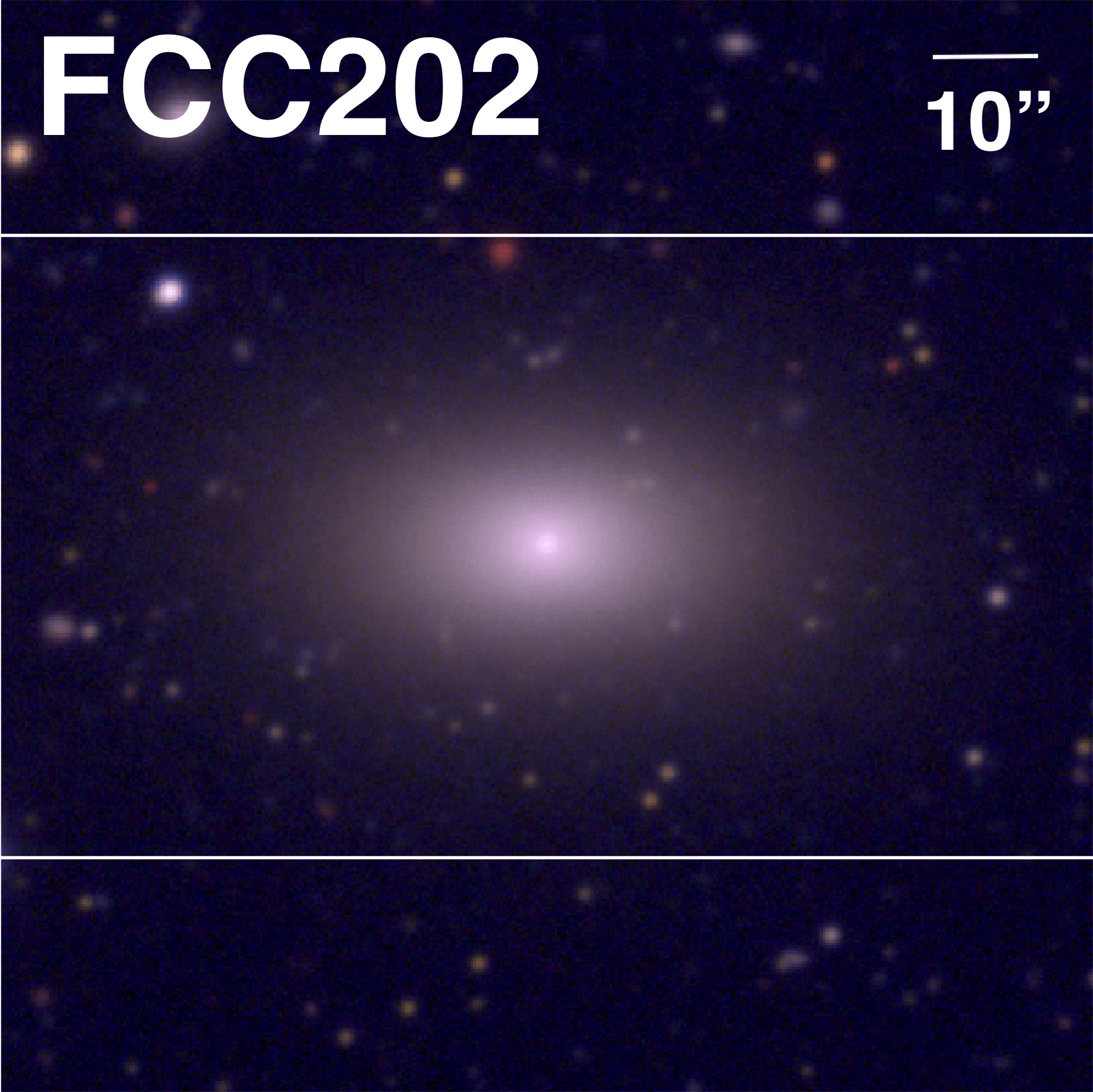}
\includegraphics[angle=0,width=0.30\linewidth]{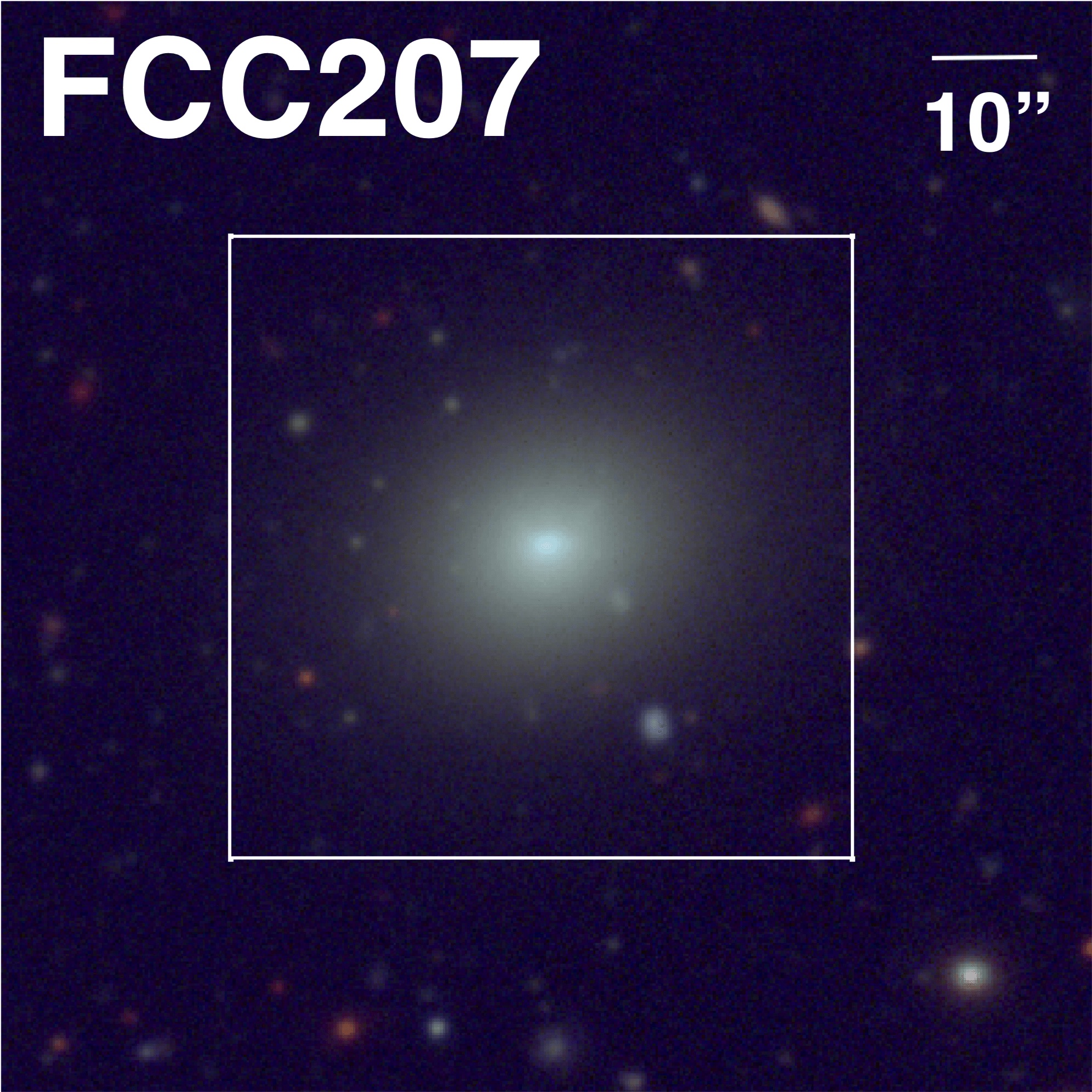}
\includegraphics[angle=0,width=0.30\linewidth]{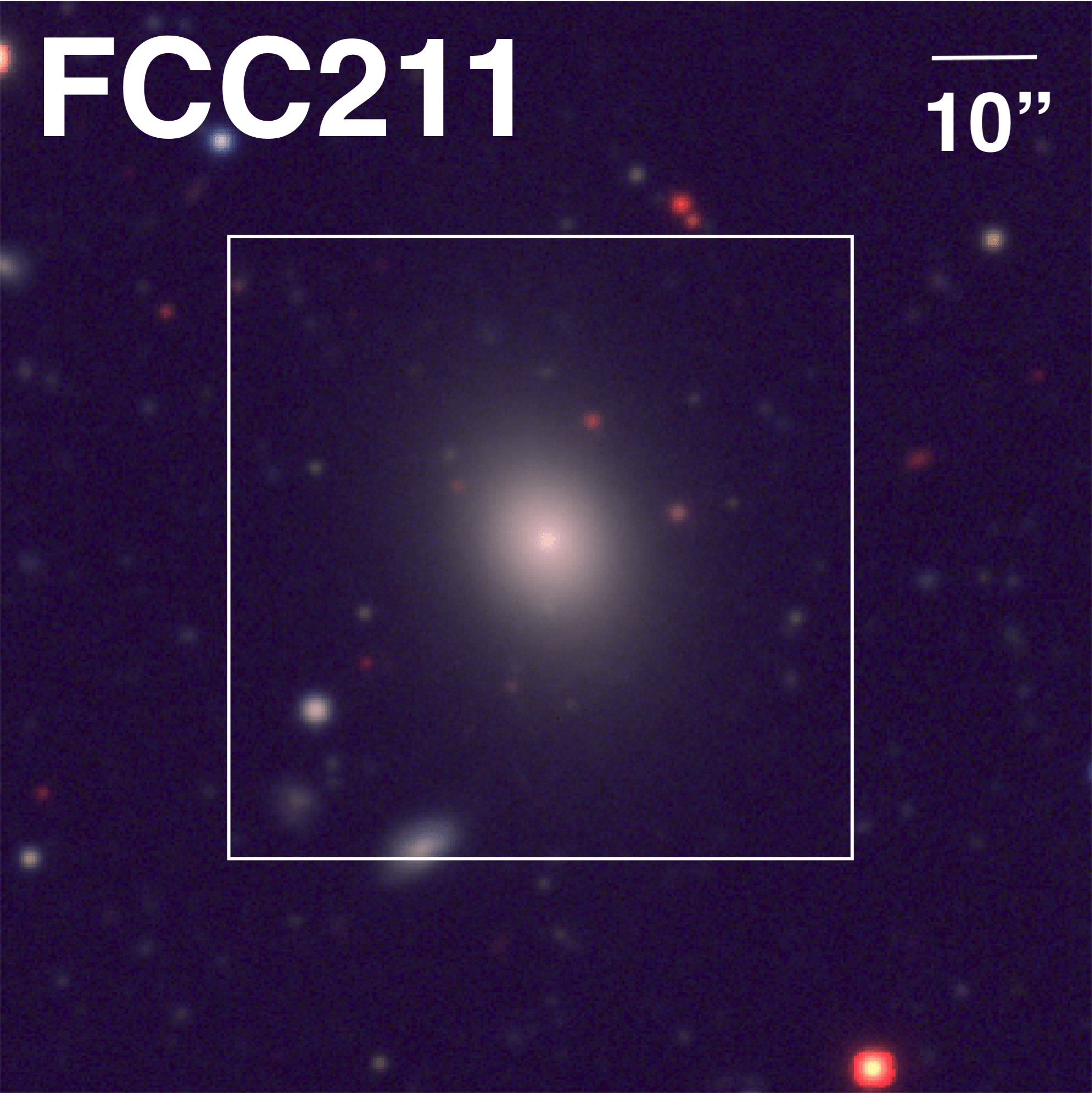}
\includegraphics[angle=0,width=0.30\linewidth]{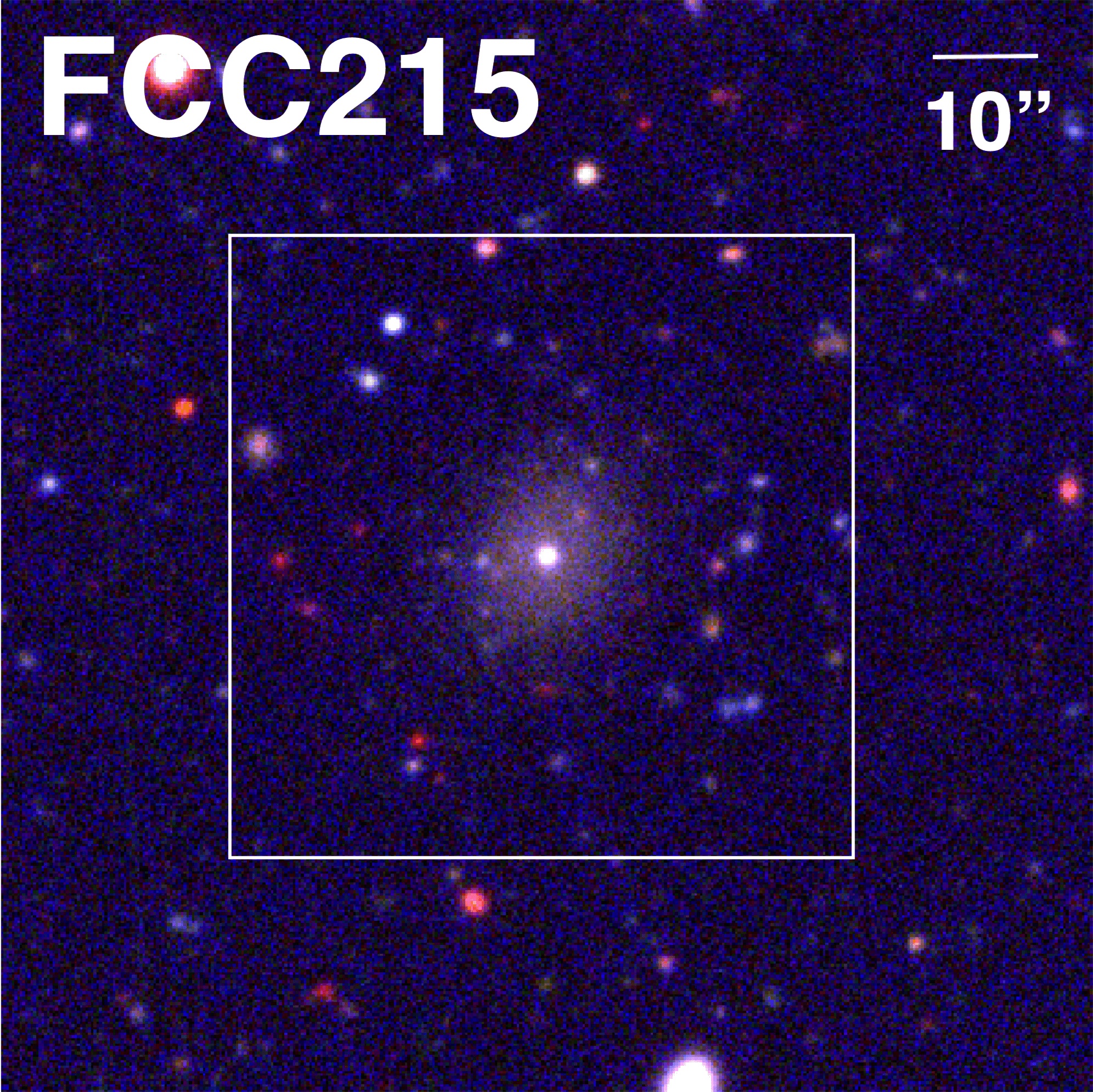}
\includegraphics[angle=0,width=0.30\linewidth]{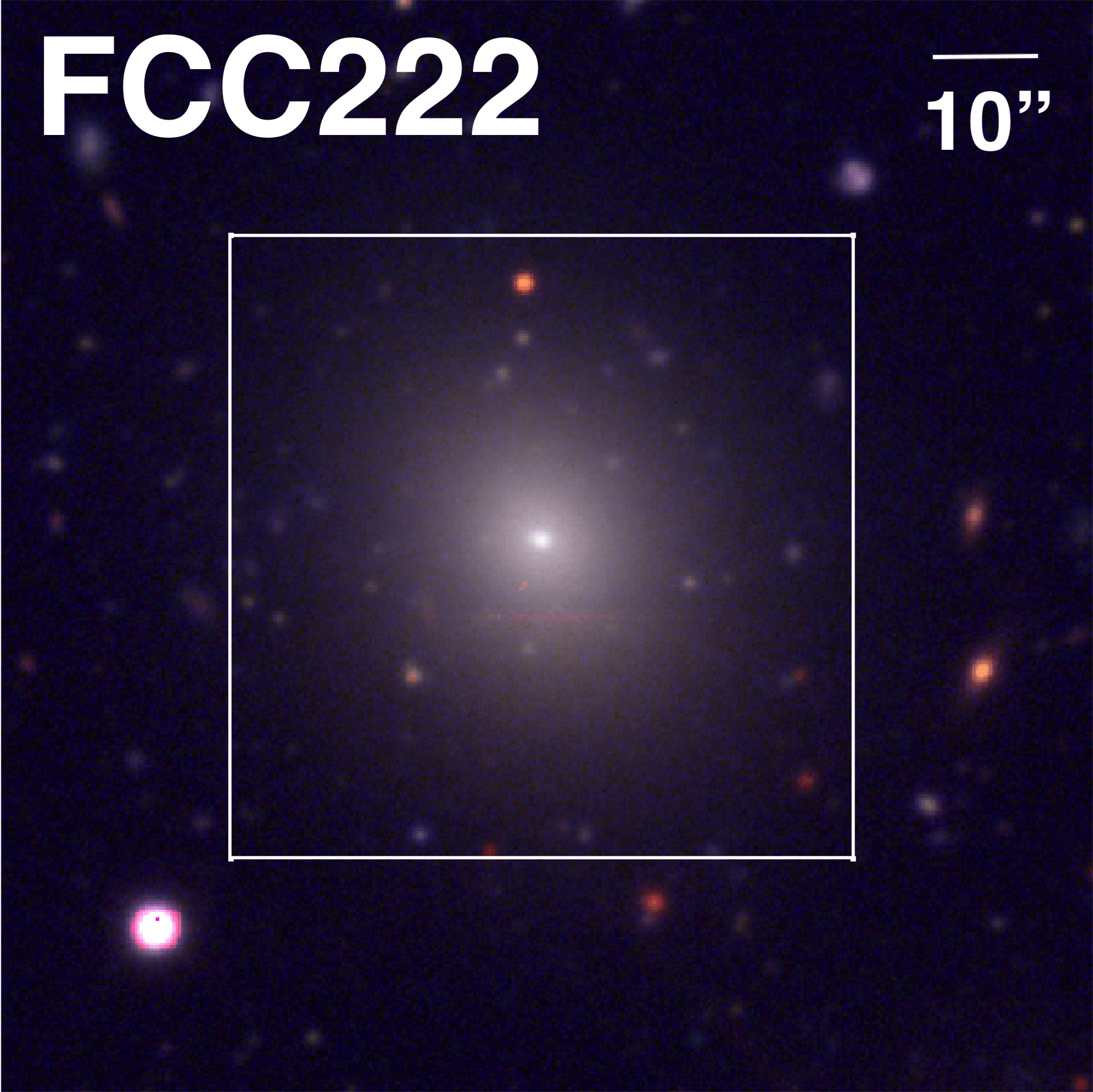}
\includegraphics[angle=0,width=0.30\linewidth]{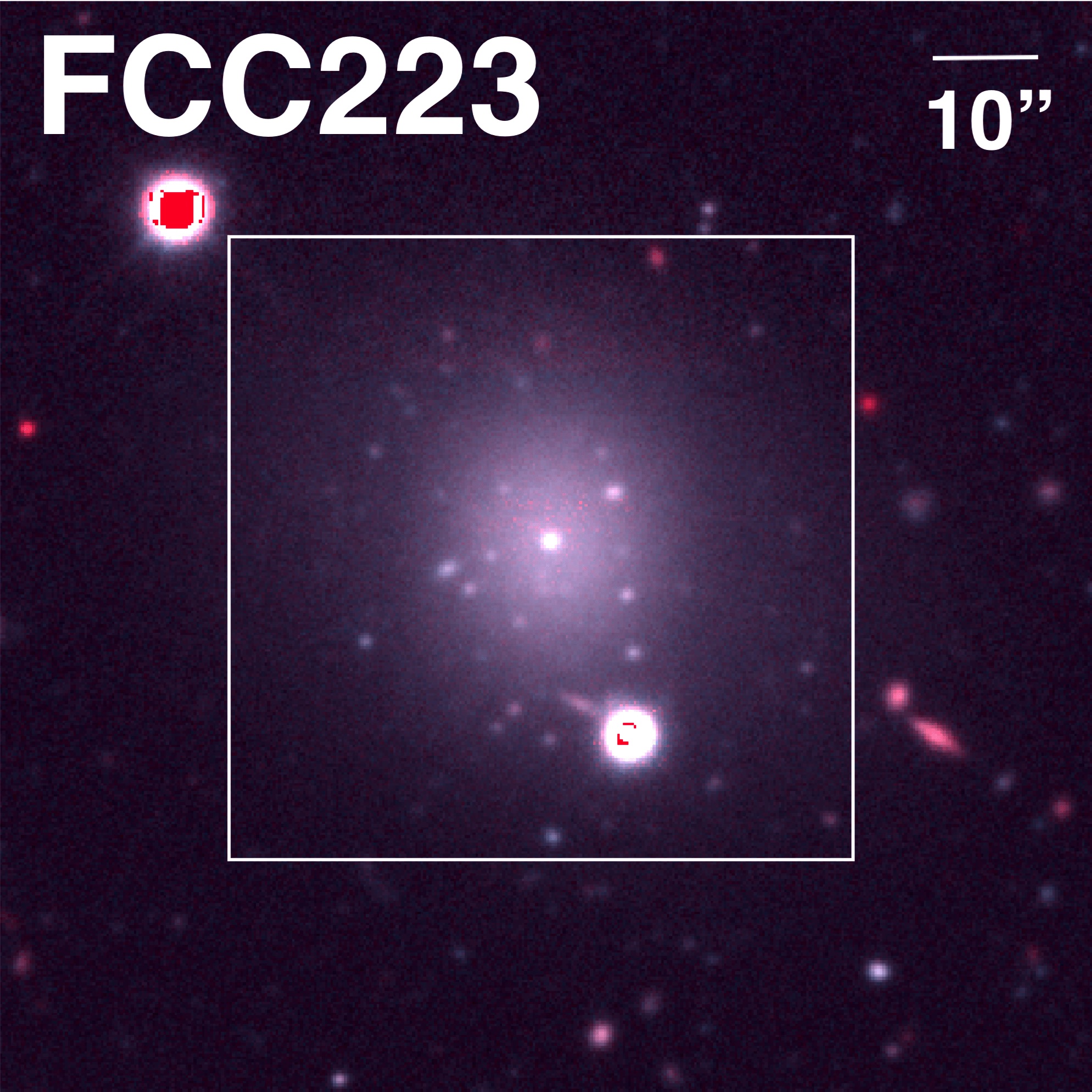}
\includegraphics[angle=0,width=0.30\linewidth]{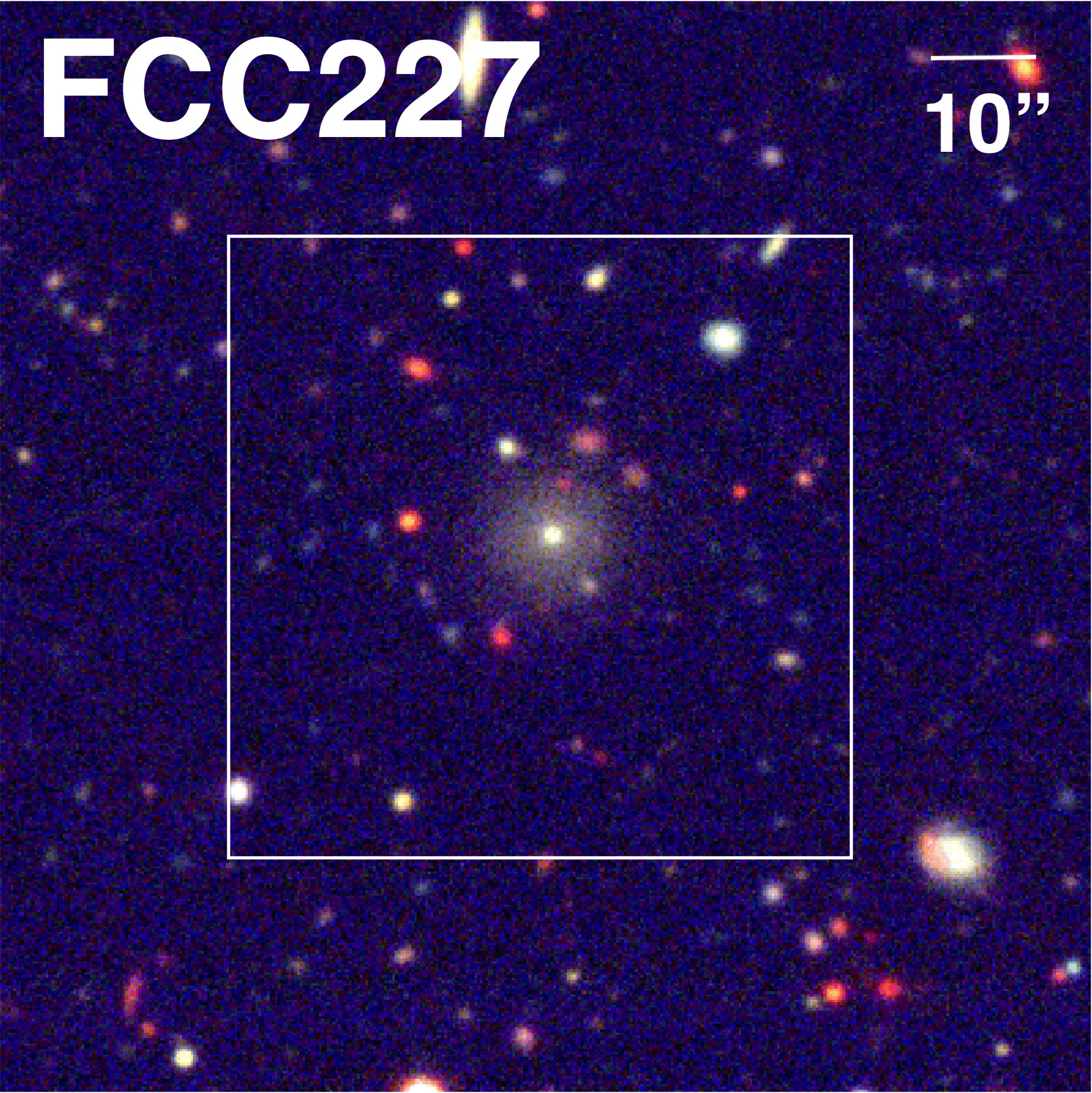}
\includegraphics[angle=0,width=0.30\linewidth]{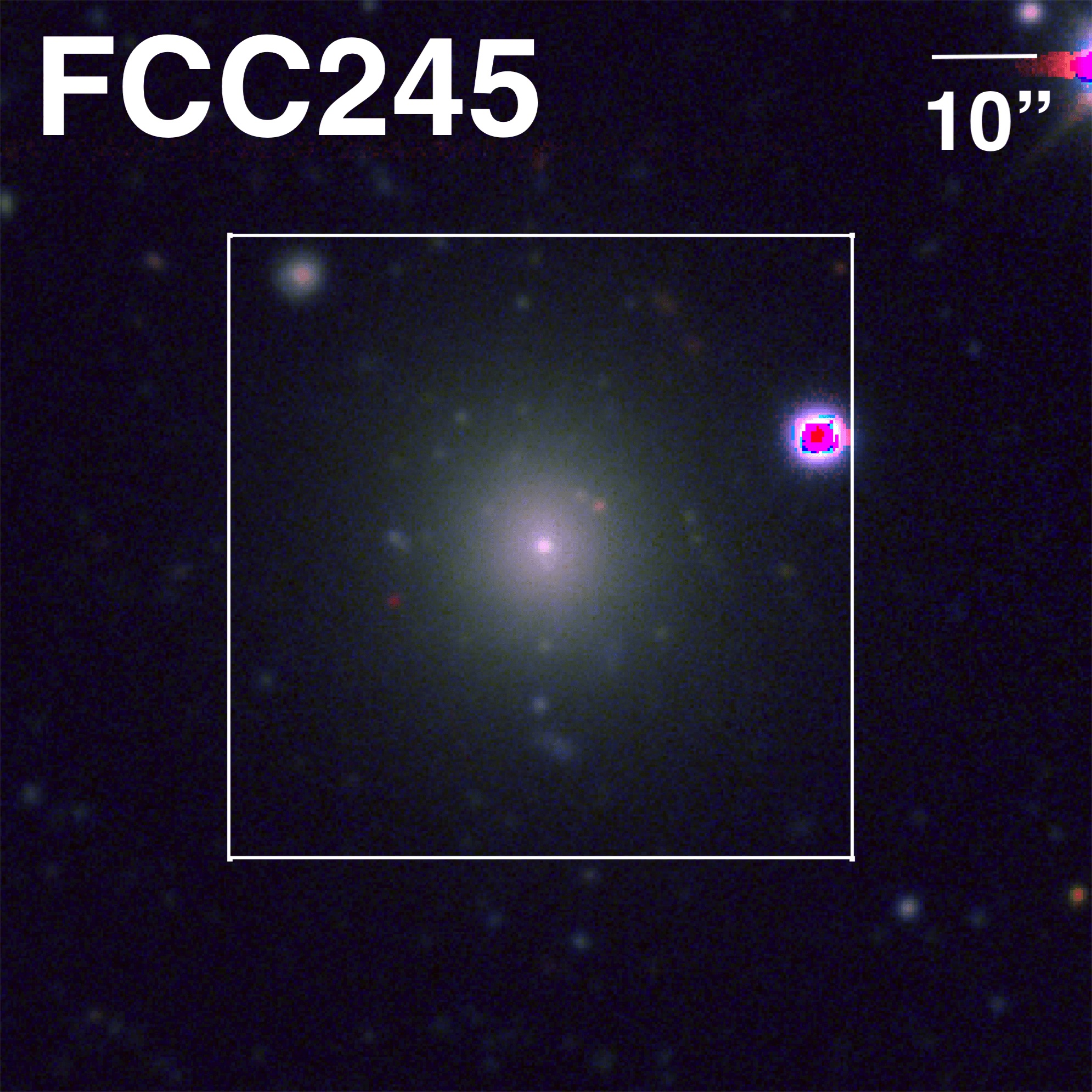}
\includegraphics[angle=0,width=0.30\linewidth]{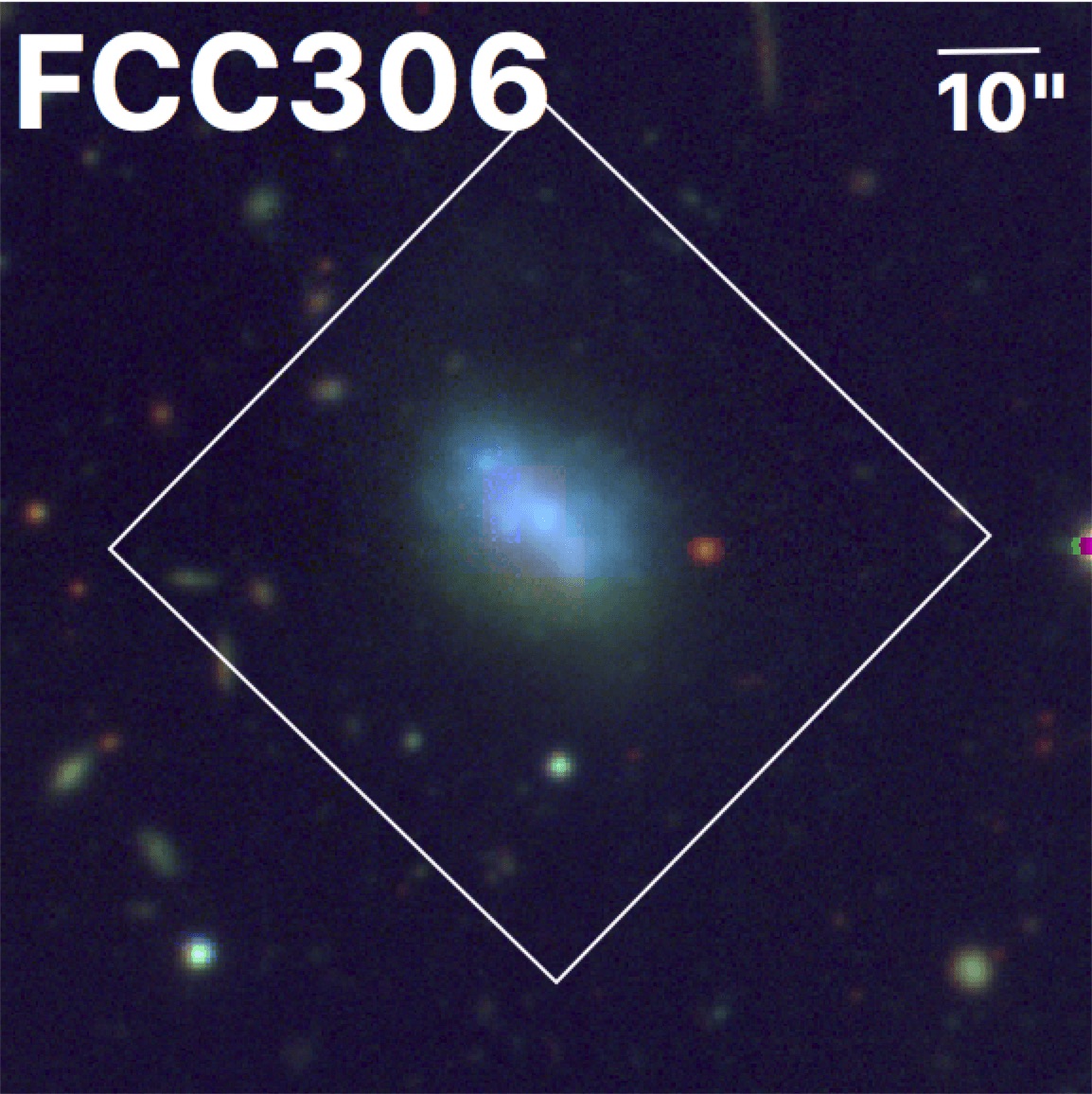}
\includegraphics[angle=0,width=0.30\linewidth]{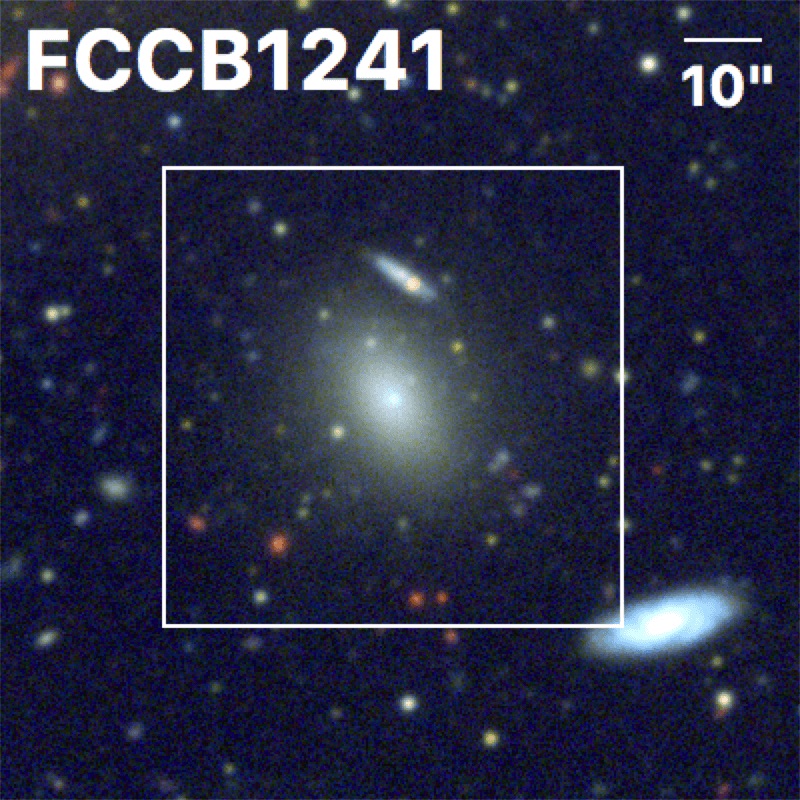}
\caption{RGB images of each galaxy in the sample, created with the NGFS $u^\prime$-, $g^\prime$-, and $i^\prime$-band images. For reference, the MUSE FOV has been marked as a white square measuring $1\arcmin\!\times\!1\arcmin$, except in the case of FCC~202, which was observed over two pointings, and the scale bar in the top right of each plot marks 10\,\arcsec, which corresponds to $\sim\!1$~kpc at the distance of Fornax. All images have been itnensity scaled to best display the nucleus, and are orientated with North upwards and East to the left.  \label{fig:NGFS_images}}
\end{center}
\end{figure}

\subsection{Imaging data}\label{sec:NGFS_data}
{The corresponding imaging data for each galaxy in this sample comes from the Next Generation Fornax Survey (NGFS), which is an ongoing multi-waveband survey of the Fornax Cluster virial sphere using Blanco/DECam for near-ultraviolet/optical and VISTA/VIRCAM for near-infrared photometry. When completed, the NGFS will provide a homogeneous panchromatic map of the Fornax  cluster in the $u^\prime g^\prime i^\prime K_s$-bands. The science goals focus on dwarf and giant galaxies, ultra-diffuse galaxies (UDGs), and compact stellar systems, such as globular clusters and ultra-compact dwarfs (UCDs), but the depth of the survey will also allow detailed studies of structures beyond the Fornax Cluster \citep[e.g.][]{Johnston_2019}. Below we provide relevant details of the NGFS data, while a more detailed overview of the NGFS observations and data reduction can be found in \citet{Eigenthaler_2018} and \cite{Ordenes_2018a}.

This study made use of the $u^\prime$, $g^\prime$ and $i^\prime$-band imaging data. The exposure times were determined to obtain a signal-to-noise ratio (S/N) of 5 at 26.5, 26.1 and 25.3 AB mag in the  $u^\prime$-, $g^\prime$- and $i^\prime$-bands, respectively. The data reduction was carried out in two steps. In the first step, the DECam Community Pipeline \citep[CP,][]{Valdes_2014}  applies the bias correction, flat-fielding and cross-talk correction. Further processing was then carried out using the \textsc{Astromatic} software, which calculated the astrometric calibration, stacked the individual images, and  applied the flux calibration using \textsc{scamp} \citep{Bertin_2006}, \textsc{swarp} \citep{Bertin_2002} and \textsc{source extractor} \citep{Bertin_1996}. The RGB images of each galaxy created from the NGFS $u^\prime$-, $g^\prime$-, and $i^\prime$-band images are shown in Fig.~\ref{fig:NGFS_images}, having been scaled to make the nuclei more visible and with the MUSE field-of-view marked as a white box.


\section{Modelling the galaxy with \textsc{buddi}}\label{sec:BUDDI}

Since the nuclei are faint compared to their host galaxies, any spectrum extracted from the region of the nucleus will contain light from both components. As a result, analysis of the nuclear region will suffer contamination from the light of the host galaxy and, without careful modelling, will  skew the results for the NSC by an unknown amount towards the properties of the rest of the galaxy. The only way to be able to extract the pure, uncontaminated spectrum of the nucleus is to use \textsc{buddi} to model both components in order to quantify and remove the contamination from the host galaxy.  While earlier attempts have been made at separating the light from different components in galaxies with long-slit spectroscopy \citep[e.g.][]{Silchenko_2012, Johnston_2012, Johnston_2014},  IFU datacubes, which provide both photometric and spectroscopic information simultaneously, allow more complicated morphologies to be modelled reliably.

A datacube can be considered as a series of narrow-band images at each wavelength step. Therefore, if a given image has sufficient signal-to-noise (S/N), conventional bulge-disc decomposition techniques, such as \textsc{Galfit} \citep{Peng_2002,Peng_2010}, can be applied to that image to derive the structural parameters. However, due to the small steps in wavelength in an IFU datacube, the image slices are more strongly affected by sky features and the S/N may be too low for a reliable fit at certain wavelengths. \textsc{GalfitM} \citep{Haeussler_2013, Vika_2013} is a modified form of \textsc{Galfit} developed by the MeGaMorph team that is able to fit multi-waveband images of a galaxy simultaneously, employing user-defined Chebyshev polynomials to constrain the variation in the structural parameters over the full wavelength range. This constraint allows \textsc{GalfitM} to use information from the entire dataset to fit the galaxy in each image, and thus allow the user to derive reasonable estimates of the structural parameters to be obtained for images with lower S/N. \textsc{buddi} takes this modelling technique one step further to apply two-dimensional light profile fitting to all the image slices in an IFU datacube, allowing the user to extract the uncontaminated spectrum of each component included in the model. Full details of how \textsc{buddi} works can be found in \citet{Johnston_2017}, and a summary of the key steps is given in the following sections.

\subsection{Step 0: Create the mask and PSF profile}\label{sec:step_0}
Several initial preparation steps must be carried out before the galaxy can be modelled. The first step is to create a bad pixel mask identifying those pixels outside of the MUSE field-of-view (FOV) within the image, and also those pixels containing light from foreground stars or background objects that might affect the fit to the galaxy. 

The second preparation is to create a point-spread function (PSF) for each image in the datacube that can be convolved with the fit to that image. Ideally the PSF datacube would be created from non-saturated stars in the science image, but the large size of the majority of the galaxies within the MUSE FOV and the lack of isolated stars in the fields meant that this option was not always possible. In such cases, the PSF datacube was created using stars in the separate sky exposures, which were observed as part of the same observing block as the galaxy exposures, or using the standard star field as a final option, where the standard star field was selected to match the seeing of the final datacube as closely as possible. The peak flux for the stars used for creating the PSF model was checked to ensure they were not saturated and were within the linearity limits of the MUSE CCDs, and in general between 1 and 6 stars were used to create the PSF model for each galaxy. While this approximation of the PSF is not perfect, it was found to provide a satisfactory fit for each galaxy.

\subsection{Step 1: Obliterate the kinematics}\label{sec:step_1}
At specific wavelengths throughout a galaxy datacube, the galaxy components will appear asymmetric due to the rotation-induced Doppler shift across the galaxy. In these cases, part of the galaxy will be fainter (brighter) than the other as the datacube shows the light there in an absorption (emission) feature while the other side of the galaxy will reflect the luminosity of the continuum. Since \textsc{Galfit} is unable to model such asymmetric structures, the first step within \textsc{buddi} is to apply the Voronoi tessellation technique of \citet{Cappellari_2003}  and then to measure the kinematics of the binned spectra with the penalised Pixel Fitting software (pPXF) of \citet{Cappellari_2004}. pPXF combines a series of stellar template spectra of known ages and metallicities, and convolves them with a range of line-of-sight velocity distributions (V$_{\rm LOS}$) and velocity dispersions ($\sigma$) to produce a model spectrum that best fits the binned galaxy spectrum and its kinematics. 
In this study, the template spectra used in the fits were those of \citet{Vazdekis_2010}, which are based on the MILES stellar library of \citet{Sanchez_2006} and consist of 156 template spectra ranging in metallicity from -1.71 to +0.22, and in age from 1 to 17.78 Gyrs. No clear rotation was detected in 11 of the galaxies, meaning that no action was needed to normalise the kinematics in those cases. The only galaxy that was found to have significant rotation was FCC~202, and it was analysed in more detail in \citet{Mentz_2016}.

\subsection{Step 2: Fit the white-light image}\label{sec:step_2}
A typical MUSE datacube has of the order 3600 image slices, and to fit all of these images simultaneously with \textsc{GalfitM} would require a huge amount of CPU time. To counter this issue, \textsc{buddi} carries out the fit using a multi-step process. The first step is to create a white-light image by stacking all the image slices within the datacube, and then to model the galaxy in this image with \textsc{GalfitM}. This step allows the user to quickly determine an approximate best-fitting model for the galaxy, including the number of components necessary for a good fit and appropriate starting parameters for the model.

Each galaxy was first modelled as a single S\'ersic profile with a PSF component to represent the nucleus, and additional S\'ersic models were added to the fit until the residual image, created by subtracting the best fit from the input image, showed the fewest artifacts with the minimum number of components. When determining the best fit, the aim was to model as much of the light from the galaxy as possible, leaving behind only faint, asymmetric structures in the residual images and thus reducing the amount of contamination from the light of the host galaxy in the model of the NSC. An example of a fit to the white light image of FCC~211 can be seen in Fig.~\ref{fig:galfit_fit}, showing the white light image, the best-fitting model, the residuals, and the models of each component. The one-dimensional light profile fit along the major axis has also been plotted for comparison. In general, all galaxies were found to be well modelled using between 1 and 4  S\'ersic profiles representing the host galaxy and a PSF profile to model the NSC.

\begin{figure}
\begin{center}
\includegraphics[angle=0,width=\linewidth]{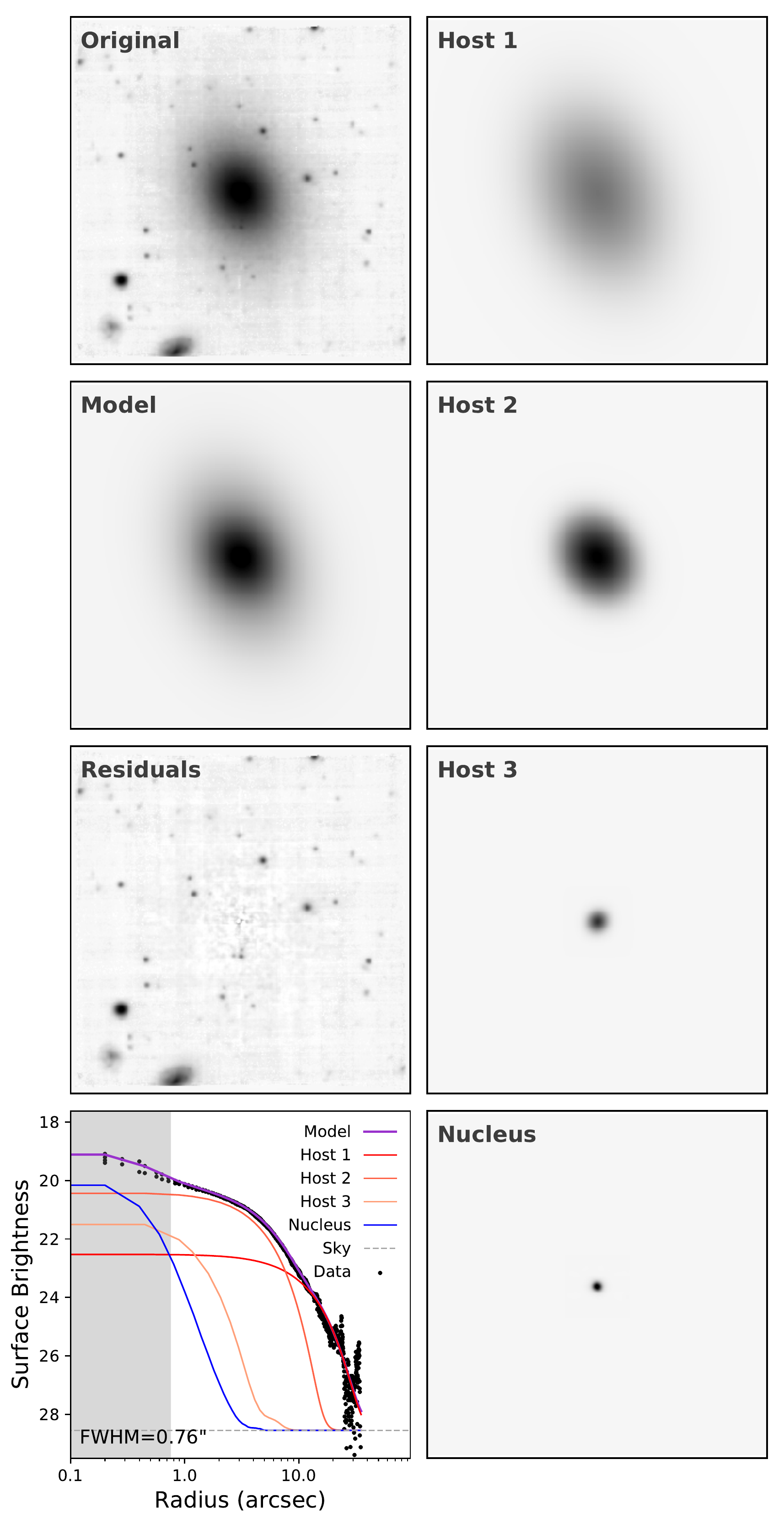}
\caption{An example of the fit to the white-light image of FCC~211 showing the input image, the best-fitting model and the residual image in the left column, and the models for each component in the right column. All images have been scaled to the same flux scale for comparison, and are orientated with North upwards and East to the left. In the bottom left plot, the light profile of the galaxy within a slit of width 1\arcsec\ along the semi-major axis has been plotted against radius as the black points, and the one-dimensional light profiles for each component and the combined model have been overplotted as the solid lines. The grey box represents the mean seeing FWHM from the final datacube.
 \label{fig:galfit_fit}}
\end{center}
\end{figure}

One interesting finding from this step was that FCC~222  was found to have a pair of point sources at the very centre.  In this case, both objects were modelled independently, as shown in Fig.~\ref{fig:galfit_fit_FCC222}, and will be discussed further in Section~\ref{sec:membership}. 

\begin{figure}
\begin{center}
\includegraphics[angle=0,width=0.94\linewidth]{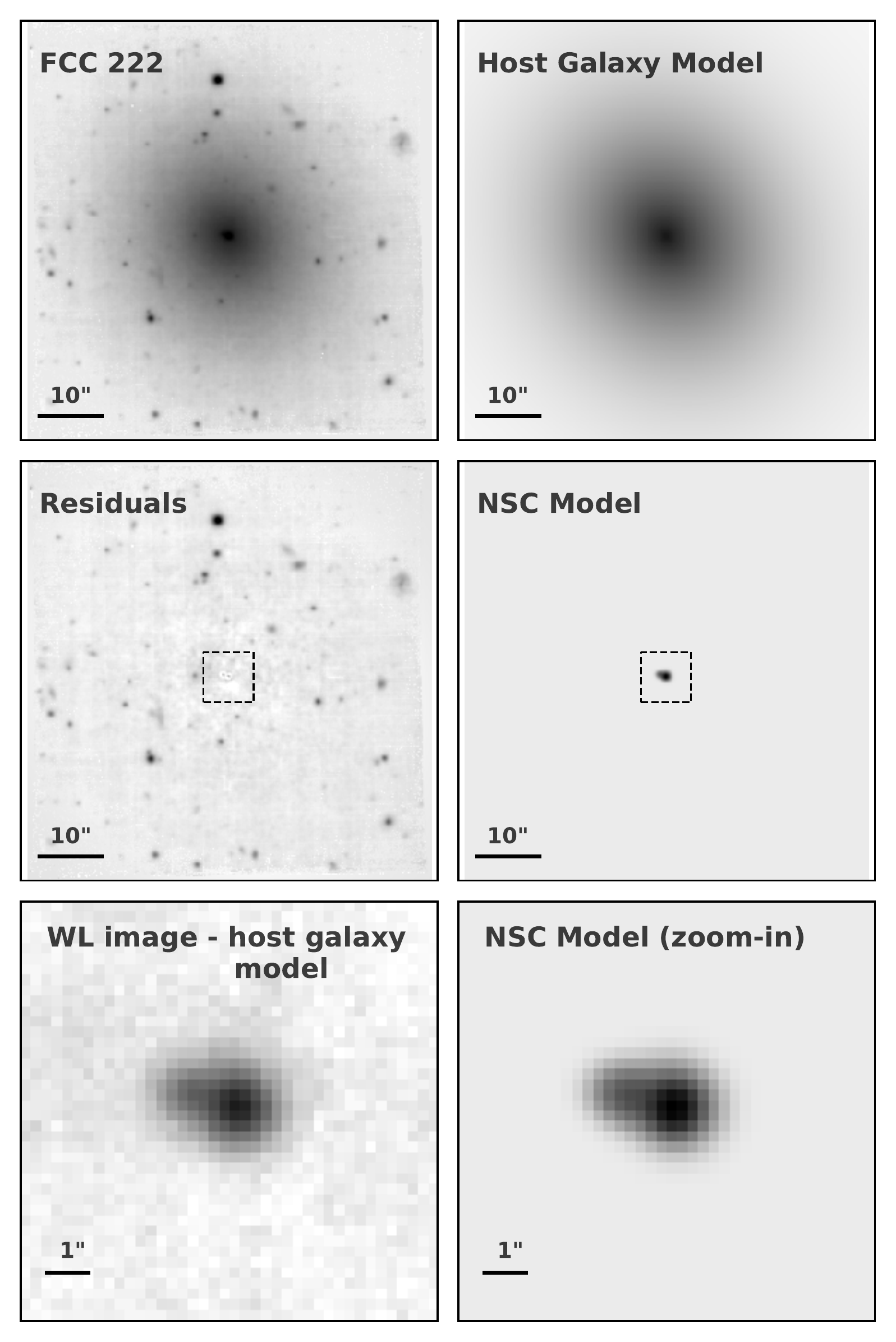}
\caption{An example of the fit to FCC~222, showing the white-light image (top-left), residual image (middle-left) and the models created for the host galaxy and the nuclear region (right-top and middle). The bottom row gives a zoom-in of the core of the galaxy, showing another residual image created by subtracting the host galaxy model from the white-light image (left) and the model created with two PSF profiles to model both  components. The field-of-view of the zoom-in regions is marked on the images above by the dashed boxes, and all images are  orientated with North upwards and East to the left. 
 \label{fig:galfit_fit_FCC222}}
\end{center}
\end{figure}

\subsection{Step 3: Fit the narrow-band images}\label{sec:step_3}
Having modelled the white-light image of the galaxy, one must next consider the effect of gradients in the structural parameters as a function of wavelength. These variations in the parameters would indicate colour gradients within  each component, which may in turn reflect the presence of stellar populations gradients \citep[see e.g.][]{Johnston_2012}.  Consequently, they should be taken into account when modeling the galaxy as a function of wavelength. 

The datacube is rebinned into a series of high S/N narrow-band images, to use the terminology of \citet{Johnston_2017}, and these images were modelled with \textsc{GalfitM} using the fit to the white light image as the initial estimates. In this work, the datacube was rebinned into 10 narrow-band images created by stacking $\sim370$~consecutive image slices, which corresponds to wavebands of $\sim350-600$~\AA\ at the blue and red ends of the spectrum respectively due to the logarithmic wavelength calibration.  This level of binning was found to be a good compromise between computing time and reflecting the variations in the parameters over the spectral range. The fit to these images was carried out multiple times for each galaxy, the first time using no constraints on the parameters to obtain a free fit, effectively simulating running \textsc{Galfit} on each image independently, and the subsequent iterations constraining the effective radius ($R_e$), S\'ersic index ($n$), axis ratio and position angle to follow a Chebychev polynomial with wavelength while the integrated magnitude was allowed complete freedom. This repetition allows the user to check that the order of the polynomial provided a suitable fit to the measurements from the unconstrained fits and that the initial estimates for the constrained fit were appropriate to create a reliable model at each wavelength. Additionally, the residual images were checked at this stage to ensure that  a good fit was achieved in all images. In all galaxies, the parameters were modelled with Chebychev polynomials of order 1 (constant with wavelength) or 2 (linear variation with wavelength).

\subsection{Step 4: Fit the image slices and extract the decomposed spectra}\label{sec:step_4}
The final step of the fitting process in \textsc{buddi} is to model the individual image slices within the datacube. The datacube was divided into batches of ten image slices for each fit with \textsc{GalfitM}, and each set of images was modelled by constraining the structural parameters to the values derived at each wavelength from the polynomials from the previous step. Only the integrated flux was allowed complete freedom to vary. 

After modelling each image slice with \textsc{buddi}, the parameters determined by \textsc{GalfitM} can be used to create the decomposed spectra for each component. One way to carry out this step is to plot the total flux for each parameter as a function of wavelength to create a one-dimensional spectrum of that structure with high S/N. Examples of the spectra for the host galaxy and NSC of FCC~222 created in this way are given in Fig.~\ref{fig:spectra}.  These spectra were used in the analysis discussed in the following sections. The S/N of the final spectra were found to depend on the depth of the observations and luminosity of the NSC. The spectra from the host galaxies were found to have a S/N in the range of $20-80$ per Angstrom, measured at $\sim6100$~\AA, while the NSCs had lower S/N of around $5-50$ per Angstrom. Cases with the lowest S/N were found to be due to shallower observations (e.g. FCCB~1241) or very faint NSCs (e.g. FCC~227). An alternative way to derive the spectra is to use \textsc{GalfitM} to create images of the structural components at each wavelength, which can then be combined to create model datacubes for the nuclei and host galaxies.

\begin{figure*}
\begin{center}
\includegraphics[angle=0,width=\linewidth]{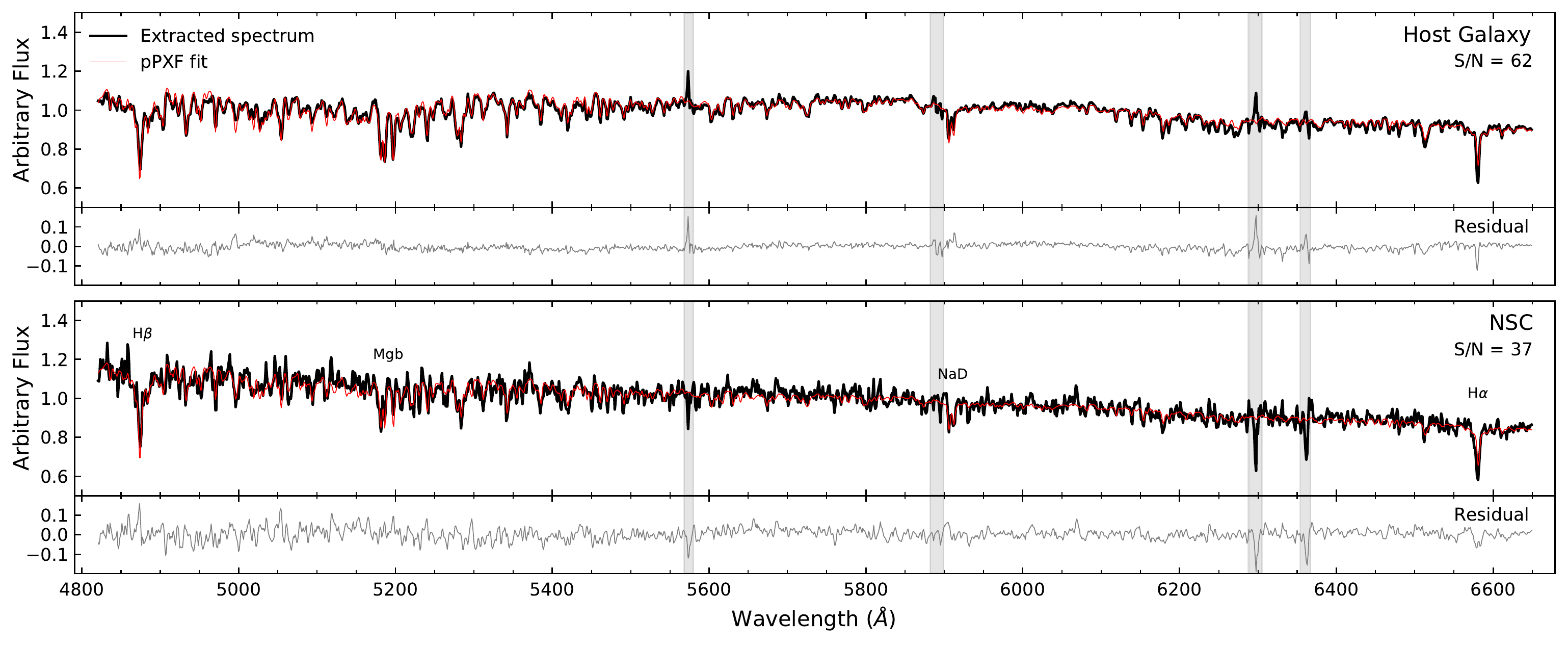}
\caption{An example of the spectra extracted for the host galaxy (top) and NSC (bottom) in FCC~222, with the best fit to the kinematics from pPXF superimposed in red. The flux for both spectra have been normalised, and the residuals are given underneath each plot. The grey boxes mark the positions of residual sky lines, and the S/N per Angstrom of each spectrum is given below the label on the right of each spectrum.
 \label{fig:spectra}}
\end{center}
\end{figure*}

The best-fitting models to the galaxies in this sample were found to consist of between 2 and 5 components, with S\'ersic profiles modelling the host galaxies and a PSF profile representing the NSCs. Studies of NSCs with the HST by \citet{Boker_2002}, \citet{Cote_2006}, \citet{Turner_2012} and \citet{Georgiev_2014} found  that the  half light radii of these structures were typically $3-5$~pc, which corresponds to an angular size of $\sim0.03\arcsec -0.05\arcsec$ at the distance of the Fornax Cluster. \citet{den_brok_2014} measured NSC sizes of up to 30~pc in dwarf galaxies in the five-times more distant Coma Cluster, which still represents an angular size of $\sim0.3$\arcsec. Consequently, throughout this paper, we assume that the NSC is unresolved and is modelled by the central PSF profile while the more extended S\'ersic profiles model the host galaxy.  This approach is similar to that of \citet{Janz_2012} in the SMAKCED project, in which $H$-band images of dwarf galaxies in Virgo were modelled with \textsc{Galfit}. The physical parameters of these different components within the host galaxy will be discussed in Section~\ref{sec:structural_params}, but a detailed analysis of the properties of these components is beyond the scope of this paper and will be addressed in a later paper. For the analysis presented in this work after Section~\ref{sec:structural_params}, the spectra from each component used to model the host galaxy will be combined to create a single spectrum representing the total integrated light from the host galaxy.

\bigskip

\noindent While the technique and results presented above show promise to better understand how faint structures such as NSCs formed and evolved, it has never before been applied to extract the spectra of such faint and compact structures within galaxies. Therefore, a final test of the reliability of this method was merited. For this test, a series of simulated datacubes were created of mock galaxies with different stellar populations, and run through \textsc{buddi} in the same way as the MUSE datacubes to check that the extracted stellar populations matched those that were used to create the models. Details of these tests and the results are given in Appendix~\ref{sec:tests}.


\section{Analysis: Physical parameters}\label{sec:analysis1}

\subsection{Structural Parameters of the Nuclei and Host Galaxies}\label{sec:structural_params}
All galaxies in this sample were found to be modelled best with between 1 and 4  S\'ersic profiles representing the host galaxy and a PSF profile as the potential NSC, where the larger galaxies were found to require more components than the smaller members of the sample. FCC~222, FCC~245, FCC~306 and FCCB~1241 each required one component to model the host galaxy, FCC~215, FCC~223 and FCC~227 required two components, and FCC~188, FCC~202 and FCC~211 required three components, and the most massive galaxy in the sample, FCC~182, required 4 components. Models with similar complexity were found for dwarf galaxies in the Virgo Cluster by \citet{Janz_2012} and \citet{Spengler_2017}.

The fit to FCC~207 is worth noting at this point. This galaxy required a double S\'ersic profile in addition to the PSF profile at the core, where the more compact S\'ersic profile resembles a flattened disc and is offset from the centre of the galaxy by $\sim 1.5$\arcsec\ ($\sim 147$\,pc at the distance of Fornax), as shown in Fig.~\ref{fig:galfit_fit_FCC207}. An inspection of the spectra of each component showed that while all three components contain gas emission, the spectrum of the PSF component shows no clear stellar continuum or absorption features, and the features in the spectrum for the flattened disc have very low S/N. A study of this galaxy by \citet{DeRijcke_2003} found that the nucleus in FCC~207 is elongated and kidney-shaped due to a region of dust obscuration to the north, however with the mean seeing in the MUSE datacube of 1.22\arcsec\ the nucleus is unresolved in this data set. No HST data have been obtained to our knowledge.~Consequently, with the data available, it is unclear whether these two central components represent a faint NSC and a compact disc, or a gaseous disc, perhaps partly obscured by dust, that is fuelling star formation and creating the NSC in situ.~Future high-resolution imaging is well encouraged.

\begin{figure}
\begin{center}
\includegraphics[angle=0,width=0.94\linewidth]{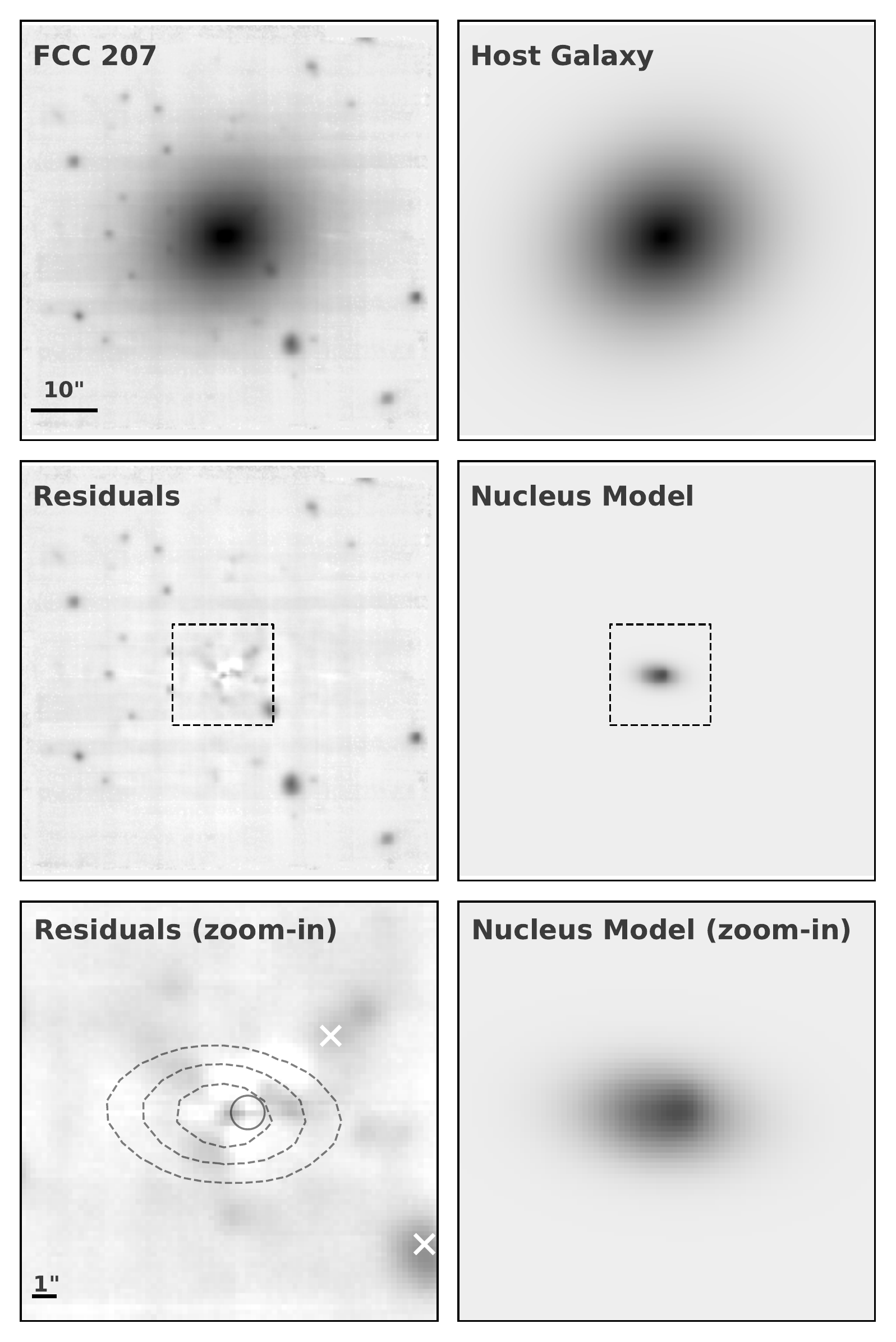}
\caption{An example of the fit to FCC~207, showing the white-light and residual images in the left column, and the models for the host galaxy and the nuclear region in the right column. The bottom row shows a zoom-in of the nuclear regions, outlined by the boxes in the middle row. In the residual zoom-in image, the circle and ellipses represent the FWHM of the PSF profile and the contours of the disc respectively, and the white crosses indicate background objects. All images are oriented with North upwards and East to the left. 
 \label{fig:galfit_fit_FCC207}}
\end{center}
\end{figure}

When fitting the narrow-band images with \textsc{buddi}, as discussed in Section~\ref{sec:step_3}, one obtains information on the structural parameters of each component as a function of wavelength. This information can reveal colour gradients within the galaxy as they appear as gradients in the size or S\'ersic index as a function of wavelength, as well as providing details on the morphology of that component \citep{Vulcani_2014,Kennedy_2015, Kennedy_2016}. 

Figure~\ref{fig:structural_params} shows the variation in the apparent magnitude as a function of wavelength for each component included in the models for the host galaxy and NSC. It can be seen that almost all of the host galaxy components are brighter at redder wavelengths, and the NSC profiles generally follow a similar  trend. Redder colours typically reflect older ages or higher metallicities within the stellar populations present. Since ten of the galaxies in this sample show no clear evidence of gas emission in the MUSE datacubes, they are likely to be quiescent galaxies, which would agree with the red colours. The central disc component of FCC~207 shows a particularly steep gradient, indicating that it is also very red. This component coincides with the emission region in that galaxy, which likely reflects ongoing, potentially dust-obscured star formation. Consequently, the colour of this structure either indicates that the star formation is occurring within an already very old stellar population, or that it is producing stars with high metallicity. An alternative explanation could be that the red colour is evidence of dust within the core of this galaxy. The residual image from the fit to this galaxy in Fig.~\ref{fig:galfit_fit_FCC207} shows some structure at the core of the galaxy, which may represent the disc of dust identified by \citet{DeRijcke_2003}. The exception to this trend is the host galaxy of FCC~306, which shows a weakly decreasing magnitude with wavelength indicating that it is bluer than the rest of the sample. Since this galaxy contains gas emission and ongoing star formation throughout its structure, this bluer colour is unsurprising.

\begin{figure}
\begin{center}
\includegraphics[angle=0,width=\linewidth]{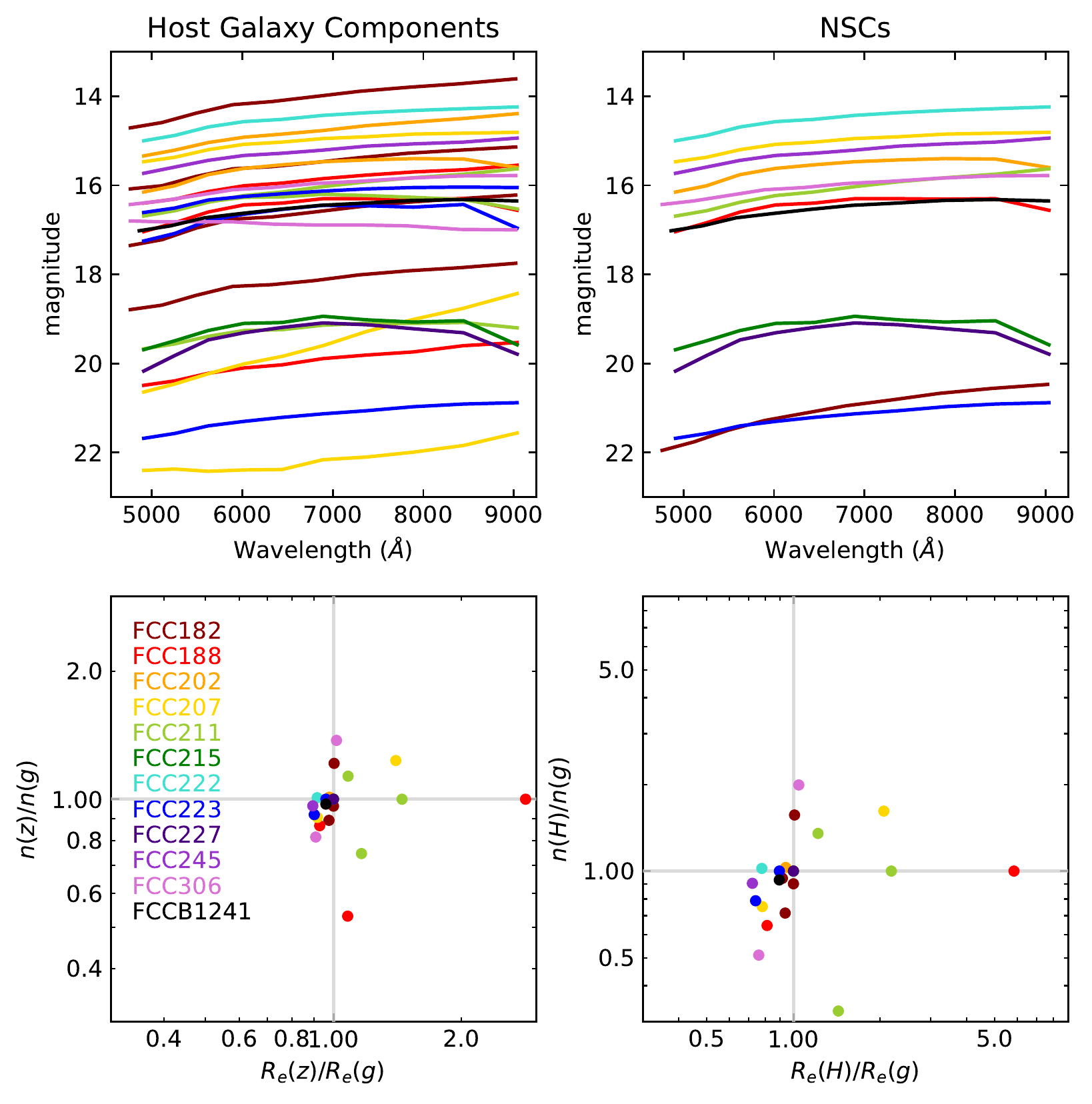}
\caption{\textit{Top}: the variation in the magnitude of each component used to model the host galaxy (left) and the NSC (right) as a function of wavelength. \textit{Bottom}: the gradients in $n$ and $R_e$ as calculated from the polynomial fits to each component at the central wavelengths of the $z$ (left) and $H$-bands (right) relative to that of the $g$-band. The colours represent each galaxy, as indicated in the bottom-left plot, and the grey horizontal and vertical lines reflect ${\cal N}=1$ and ${\cal R}=1$, respectively. 
 \label{fig:structural_params}}
\end{center}
\end{figure}

The gradients in the effective radius and S\'ersic index of each component are also given in Fig.~\ref{fig:structural_params}. In order to better display these gradients, we have adopted the notation ${\cal R}^{x}_{g} = R_e(x)/R_e(g)$ and ${\cal N}^{x}_{g} = n(x)/n(g)$ respectively, where $x$ is the selected waveband to be compared relative to the $g$-band. Physically, ${\cal R}$ gives the variation in the size of the galaxy as a function of wavelength, which can translate into a colour gradient, while ${\cal N}$ is an indicator of how the galaxy profile shape and central concentration of light varies as a function of wavelength. ${\cal R} < 1$ reflects that the galaxy is larger in the blue end of the spectrum, while ${\cal N} > 1$ indicates higher S\'ersic indices at redder wavelengths due to higher concentrations of red light in their centres \citep{Vulcani_2014, Kennedy_2015}. Furthermore, \citet{Kennedy_2016} found that when modelling multiple components within the same galaxy, a high value for ${\cal N}$ can indicate the presence of a substantial disc within that component.

\citet{Kennedy_2015,Kennedy_2016} used \textsc{GalfitM} to measure these gradients in multi-waveband imaging data of  massive galaxies in the GAMA survey to calculate the gradients in ${\cal R}$ and ${\cal N}$ for the $H$-band relative to the $g$-band for their samples of more massive galaxies, while \citet{Vulcani_2014} plotted the results for the $ugrizJHK$ bands relative to the $g$-band measurements. The polynomial fits to the MUSE narrow-band images described in Section~\ref{sec:step_3} allows one to derive reliable estimates for ${\cal R}^{z}_{g}$ and ${\cal N}^{z}_{g}$ using the central wavelengths of these wavebands as they lie within the MUSE wavelength range. Estimates for ${\cal R}^{H}_{g}$ and ${\cal N}^{H}_{g}$ can also be calculated through extrapolation if one assumes that the polynomials are reliable outside of the MUSE spectral range. Figure~\ref{fig:structural_params} plots the gradients in the parameters for the host galaxy components in both the $H$ and $z$-bands relative to the $g$ band to allow a direct comparison to \citet{Vulcani_2014} and \citet{Kennedy_2015, Kennedy_2016}. Note that only the magnitudes are given for the NSC components since they were modelled as PSF profiles.

It can be seen that the gradients for ${\cal R}$ and ${\cal N}$ for the components used to model the host galaxies are mostly close to unity, indicating little or no variation with wavelength. Such a result is unsurprising given that \citet{Kennedy_2015} found that ${\cal R}^{H}_{g}$ is generally close to unity for fainter and low-$n$ galaxies. Furthermore, the low values for ${\cal N}$ reflect an absence of rotating discs within these galaxies, which is in agreement with the flat velocity maps measured in the early step of \textsc{buddi}.

One exception to this trend is the innermost component of FCC~188, which shows ${\cal R} > 1$ and ${\cal N} = 1$. While this effect may simply be magnified artificially during the normalisation process due to the compact size of this component, it could also reflect a colour gradient within that component, such that the bluer light is more centrally concentrated than the red light. Additionally, the inner component within FCC~207 can be seen to have relatively high values for both ${\cal R}$ and ${\cal N}$, which likely reflects the disky nature of this component and a colour gradient created through stellar populations gradients or by the presence of dust. 

An alternative explanation for these trends is that the wavelength-dependent models created by \textsc{buddi} are instead fitting the different distributions of distinct stellar populations within the host galaxy, which in turn can tell us about how the host galaxy has evolved. However, a detailed analysis of the properties of the host galaxies in these systems is beyond the scope of this paper, and will be further investigated in a future study of both nucleated and non-nucleated dwarf galaxies, and in terms of the dwarf galaxy alignment seen in the Fornax Cluster by \citet{Rong_2019}.

\subsection{Looking for evidence of mergers}\label{sec:mergers}
After completing a fit to a galaxy datacube, \textsc{buddi} creates a residual datacube by subtracting the best-fitting model datacube from the original datacube. These residual datacubes and their white-light images were inspected after the fit to each galaxy to ensure that the best initial parameters were selected, and to look for evidence of faint structures within the galaxy that were otherwise too faint to see in the original datacube. These cubes revealed numerous globular clusters and background galaxies, but none of the datacubes displayed  evidence of shells or tidal tails, which are characteristic signatures of recent merger activity \citep{Paudel_2011}. The residual images created by fitting the deeper NGFS $g$ and $i$-band images were also inspected, and again no clear shells or tidal tails were detected. Therefore, if these  galaxies have undergone mergers or external mass accretion, it was long enough ago that the galaxy light profiles have become smooth again. However, it is possible that the multiple components used to model the host galaxy in the more massive galaxies could represent physical distributions of different stellar populations. This distribution in the stellar populations could be the result of mergers long ago and in which the orbits of the accreted or newly formed stars have now settled and aligned with the rest of the galaxy, or through multiple epochs of bursty star formation.

The core region of FCC~207  showed several features in the residual image, as can be seen in Fig.~\ref{fig:galfit_fit_FCC207}. A background galaxy lies to the northwest of the NSC at $z\sim1.06$, and a dust-obscured region exists to the north of the NSC and the flattened disc, as previously identified by \citet{DeRijcke_2003}. Several faint sources of light can also be seen within the region of the flattened disc itself, and extended structures lie to the east of this disc. With the low S/N of the spectra extracted from these regions, it has not been possible to measure the redshift of these features to determine whether they lie at the core of the galaxy or are background objects. If they are within the galaxy, they may be evidence of tidal tails resulting from a recent merger with an infalling galaxy.

\subsection{Confirmation of Galaxy Membership}\label{sec:membership}
The potential NSCs in this sample of galaxies have so far been identified as the point sources located close to the centre of the galaxy. However, with photometric data alone, it is difficult to definitively confirm whether they are true NSCs or are simply bright GCs that happen to align with the centre of the galaxy along the line of sight. \citet{Neumayer_2011}  found that NSCs occupy the kinematic core of their host galaxies, and so with the clean spectra extracted by \textsc{buddi} for the NSCs and their host galaxies, the redshifts of each component can be measured independently and compared to determine whether the central objects are truly at the cores of their galaxies.

The kinematics of each component were measured with pPXF on the spectra extracted by \textsc{buddi}, using the same methodology as outlined in Section~\ref{sec:step_1}. Examples of the fits by pPXF to the spectra extracted for the host galaxy and NSC in FCC~222 are shown in Fig.~\ref{fig:spectra}. In the galaxies that were modelled with multiple components representing the host galaxy, the spectra of each of these components were combined to give a single spectrum representing that galaxy. The differences in the line-of-sight velocities between each host galaxy and its potential NSC are plotted against the host galaxy velocity in Fig.~\ref{fig:velocities}, with the coloured bands indicating the velocity dispersion centred on the $\Delta V = 0$ line. The combined uncertainty in the velocities of the NSC and host galaxy are given as the grey error bars, and were calculated as the standard deviation in the measurements from a series of Monte Carlo simulations on the model spectrum created by pPXF with simulated levels of noise to match the S/N of the original spectrum. For the NSCs in FCC~207 and FCC~306, the emission line kinematics  were used because the stellar spectrum had too low S/N to reliably measure the kinematics. It can be seen that eleven of the potential NSCs all lie within $1-2\sigma$ of the redshift of the corresponding host galaxy, and thus these point sources have been spectroscopically confirmed to lie at the cores of their host galaxies, and are thus true NSCs. The exception is FCC~306, which lies around 3$\sigma$ from the velocity measured for the host galaxy. While one outlier in a sample of 12 is statistically expected, this galaxy was observed to show extended, uneven gas emission across the full extent of the galaxy, likely indicating knots of  star-formation activity. The knot that was identified as the potential NSC was the one closest to the photometric centre of the galaxy, as measured by the fit with \textsc{GalfitM} to the extended component. However, the velocity measurement shows that it does not lie in the kinematic centre of the galaxy, thus indicating that this potential NSC has likely been misclassified.

A second bright point source was identified close to the core of FCC~222 (see Fig.~\ref{fig:galfit_fit_FCC222}). Similar features have been seen before by \citet{Lauer_1996}, \citet{Kormendy_1999} and \citet{debattista_2006}, and possible explanations include dust absorption, a nuclear disc surrounding a supermassive black hole (SMBH), a pair of NSCs, or an NSC and foreground GC. Since the double-nucleus appearance can be seen at all wavelengths in the MUSE datacube, the effect of dust absorption can be ruled out \citep{Lauer_1996}. 

In the nuclear disc scenario, the accretion disc surrounding the SMBH takes on the appearance of a pair of bright sources due to the eccentricity of the orbits. The centre of mass of the system lies close to the fainter source and the bright source is the apoapsis, the point in the orbit furthest from the centre of mass, which is brighter because the stars linger near this point in their orbits. The presence of a nuclear stellar disc can be confirmed through spectral signatures such as rotation and an asymmetric velocity dispersion profile within the nuclear disc. However, the spectral resolution of the MUSE data prohibits a comparison of the velocity dispersions of the two point sources or an accurate measurement of the variation in the velocity across the nuclear region. Furthermore, the brighter point source lies at the photometric centre of the fit to the host galaxy, with a separation of $<0.1$\arcsec\ ($\sim10$~pc), while the fainter point source has a separation of $\sim0.94$\arcsec. While these trends cannot completely rule out the possibility for the nuclear region of FCC~222 containing an accretion disc surrounding an SMBH, they do make this scenario appear less likely. 

In order to distinguish between the final two scenarios, we look again at the kinematics. The velocity of the fainter point source has been plotted in Fig.~\ref{fig:velocities} as the hollow circle, and it can be seen that while the velocity of this object is still within the velocity dispersion range of the host galaxy, the velocity difference relative to the host galaxy is significantly larger than that of the brighter source, the NSC. Therefore, while this object may still be an infalling GC that has been detected close to its maximum separation from the NSC on sky, the more likely explanation is that it is simply a bright GC within FCC~222 which happens to be projected along the same line of sight as the NSC. Irrespective of this discussion, this result displays the strength of using \textsc{buddi} to cleanly extract and measure the redshifts of similar objects within a galaxy to  map out their positions and velocities within the system.

\begin{figure}
\begin{center}
\includegraphics[angle=0,width=\linewidth]{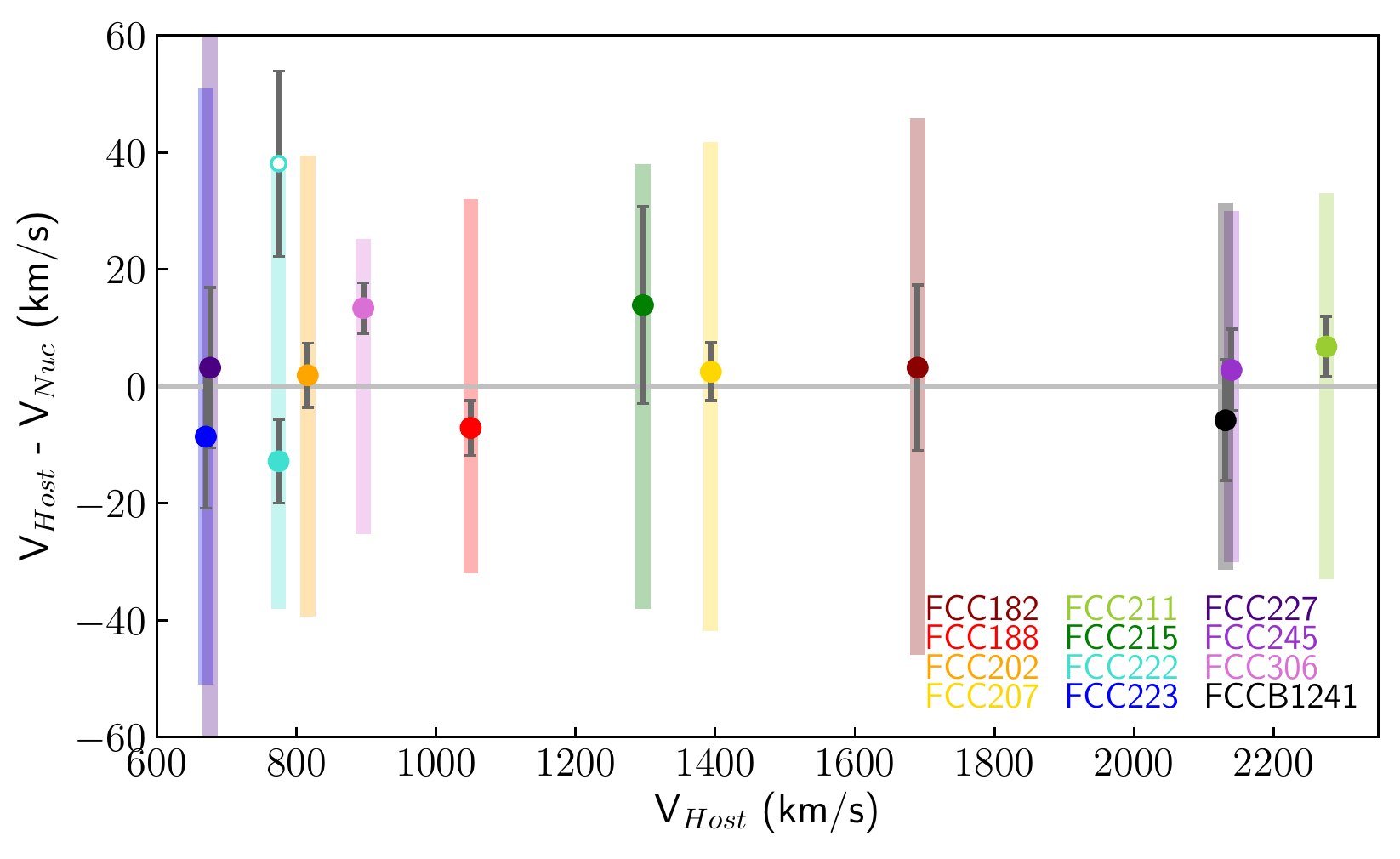}
\caption{The differences in the velocities measured from the extracted spectra for the host galaxies and their NSCs, plotted against the line-of-sight velocities of the host galaxies. The coloured bars reflect the velocity dispersion in the host galaxy spectrum, centred on the 0 velocity difference line, and the grey error bars represent the combined velocity uncertainties in the spectra of the NSC and host galaxy. The hollow circle represents the measurement from the bright GC seen in FCC~222.  \label{fig:velocities}}
\end{center}
\end{figure}

\subsection{Masses of hosts and nuclei}\label{sec:masses}
As the first step towards measuring the masses of each galaxy and their nuclei, light profile fits were carried out on the NGFS $g^{\prime}$ and $i^{\prime}$-band images of each galaxy with \textsc{Galfit} \citep{Peng_2010}, using the same starting parameters for each galaxy as the \textsc{GalfitM} fits to the MUSE white-light images. The NGFS data were selected for this analysis over the MUSE data due to the uniformity in the depth of the images and the wider field of view, allowing \textsc{Galfit} to model the fit right out to the wings of each galaxy, and the cleaner PSF profile, which gives a better fit to the NSCs. In all cases, the parameters for the component models required to fit the NGFS images of each galaxy were found to be consistent with those derived from the fits to the MUSE images.

The luminosity-weighted stellar masses were derived by fitting stellar population synthesis models to the integrated $g^{\prime}$ and $i^{\prime}$-band magnitudes of each component. The SSP models of \citet{Bruzual_2003} were used with the IMF of \citet{Kroupa_2001}, covering a metallicity range of $0.0001 \le Z/Z_{\odot} \le 0.5$ and using the luminosity-weighted ages derived for each component from its decomposed model spectrum (see Section~\ref{sec:line_strengths1}). In the three galaxies where the S/N of the extracted nucleus spectrum was too low to obtain an age measurement, the mass was calculated using the age of the host galaxy to give an upper limit on the mass. The $V$-band mass-to-light ratio was used in conjunction with the $g^{\prime}-i^{\prime}$ colours since it is the optimal luminescent passband for estimating the wavelength-dependent mass-to-light ratio from single colours, and reduced the systematic dependence on the dust extinction, metallicity and star-formation history \citep[see][for details]{Zhang_2017}. The derived masses of each component are presented in Table~\ref{tab:masses}, along with the corresponding integrated $g^{\prime}$ and $i^{\prime}$-band magnitudes and the mass fraction of the nucleus within each galaxy. As expected from the sample selection criteria, all the galaxies fall into the dwarf mass regime with masses $M<10^9 M_{\odot}$, and with the NSC mass fraction generally in the range of $\sim0.0005-0.03$.  These mass fractions are consistent with those measured by \citet{Ordenes_2018b} for NSCs in the core of the Fornax Cluster, with the one exception of the NSC of FCC215, which has a mass fraction of $\sim0.12$. However, in that case, the mass of the NSC is an upper limit derived using the age of the host galaxy because the NSC spectrum had too low S/N for an accurate age estimate. Additionally, this galaxy is one of the lowest-mass galaxies in the sample, and the mass fraction is consistent with the results of \citet{SanchezJanssen_2019}, in which it was found that the mass fractions of  NSCs are higher in lower-mass galaxies.

\begin{table*}
	\centering
	\caption{Mass estimates for host galaxies and their nuclei. }
	\label{tab:masses}
	\begin{tabular}{lrrrrrrrr} 
		\hline
		Galaxy &   & $m_{g, {\rm host}}$ & $m_{i, {\rm host}}$ & $m_{g, {\rm nuc}}$ & $m_{i, {\rm nuc}}$ & log($M_{*,{\rm host}}$) & log($M_{*,{\rm nuc}}$) & $\frac{M_{*,{\rm nuc}}}{M_{*,{\rm total}}}$  \\
		\hline
FCC~182 	& &	$14.54\pm0.03$	&	$13.64\pm0.11$	& 	$22.03\pm0.01$	&	$21.53\pm0.01$	& 	$8.95^{+0.12}_{-0.12}$		& $5.61^{+0.08}_{-0.05}$  	&	0.0005  \\
FCC~188 	& &	$16.24\pm0.03$	&	$15.13\pm0.11$	& 	$20.63\pm0.01$	&	$19.97\pm0.01$	& 	$8.89^{+0.12}_{-0.11}$		& $6.846^{+0.008}_{-0.004}$  	&	0.0090  \\
FCC~202 	& & 	$15.38\pm0.06$	&	$14.14\pm0.04$	&	$20.35\pm0.07$	&	$19.59\pm0.08$	& 	$8.95^{+0.04}_{-0.04}$		& $6.96^{+0.04}_{-0.03}$   	&	0.0101   \\
FCC~207 	& &	$16.01\pm0.01$	&	$15.03\pm0.01$	&	$21.98\pm0.03$	&	$20.87\pm0.02$	& 	$8.62^{+0.007}_{-0.007}$	& $6.06^{+0.02}_{-0.03}$   	&	0.0027   \\
FCC~211 	& &	$16.40\pm0.08$	&	$15.48\pm0.16$	&	$21.03\pm0.20$	&	$12.41\pm0.50$	& 	$8.52^{+0.16}_{-0.11}$		& $6.7^{+0.7}_{-0.7}$   	&	0.0149   \\
FCC~215 	& &	$19.99\pm0.02$	&	$19.18\pm0.07$	&	$22.40\pm0.02$	&	$21.52\pm0.06$	& 	$6.79^{+0.12}_{-0.09}$		& $5.94^{+0.05*}_{-0.10}$   	&	0.1238   \\
FCC~222 	& &	$15.85\pm0.29$	&	$14.91\pm0.53$	&	$21.23\pm0.13$	&	$20.33\pm0.23$	& 	$8.8^{+0.4}_{-0.2}$ 		& $6.45^{+0.19}_{-0.22}$   	&	0.0044   \\
FCC~223 	& &	$16.42\pm0.01$	&	$15.34\pm0.01$	& 	$21.94\pm0.02$	&	$21.09\pm0.02$	& 	$8.779^{+0.016}_{-0.013}$ 	& $6.377^{+0.010}_{-0.010}$	&	0.0039   \\
FCC~227 	& &	$20.09\pm0.05$	&	$19.14\pm0.22$	& 	$23.02\pm0.02$	&	$22.41\pm0.02$	&	$6.73^{+0.14}_{-0.17}$ 		& $5.24^{+0.02*}_{-0.03}$   	&	0.0313   \\
FCC~245	& &	$16.15\pm0.01$	&	$15.06\pm0.05$	&	$21.42\pm0.01$	&	$20.60\pm0.01$	&	$8.77^{+0.05}_{-0.04}$ 		& $6.049^{+0.010}_{-0.017}$   	&	0.0019   \\
FCC~306	& &	$15.85\pm0.03$	&	$15.50\pm0.04$	&	$23.36\pm0.05$	&	$21.51\pm0.05$		&	$7.92^{+0.02}_{-0.02}$ 		& $6.13^{+0.02}_{-0.02}$   	&	0.0160   \\
FCCB~1241	& &	$17.37\pm0.01$	&	$16.34\pm0.01$	&	$23.53\pm0.02$&	$22.70\pm0.03$	&	$8.134^{+0.013}_{-0.014}$ 	& $5.48^{+0.04*}_{-0.03}$   	&	0.0022   \\
		\hline
\multicolumn{9}{p{5.9in}}{FCC~202 and FCCB~1241 were observed to have a pair of nuclei. The mass of the central nucleus only is included in this table, but both masses are included in the mass fraction calculations. The nuclei masses with an asterisk ($^*$) mark those galaxies whose ages could not be estimated from the spectra, and so the age of the host galaxy was adopted to give an upper limit on the mass.}
	\end{tabular}
\end{table*}


\section{Analysis: Luminosity-weighted stellar populations}\label{sec:analysis2}
\citet{Ordenes_2018b} demonstrated that the masses of the nuclei of dwarf galaxies within the central region of the Fornax cluster ($\le R_{\rm vir}/4$) show a bimodality, with mean masses of $\text{log}(M/M_{\odot})\approx5.4$ and $6.3$. This bimodality also appeared to be linked to the stellar population properties of the nuclei, such that more massive nuclei tend to be more metal-poor and with a wider range in ages than the less-massive nuclei, which show a wider range in metallicity and are generally younger. 

With a median mass of $\text{log}(M/M_{\odot})\approx6.1$, the nuclei in this sample of dwarf galaxies lie in the upper mass regime of this bimodality and thus would be expected to be metal-poor. The following sections will study the stellar populations of these galaxies and nuclei to compare them to the findings of \citet{Ordenes_2018b}.


\subsection{Ages and Metallicities}\label{sec:line_strengths1}

As a first step towards understanding the origins of the nuclei, we studied the luminosity-weighted stellar populations through analysis of the line strengths. Since the light from a galaxy is dominated by the youngest, brightest stars present, the ages and metallicities derived through this analysis reflect the most recent star-formation activity, and thus give an estimate of when the star formation ceased and the enrichment of the gas that fuelled it. 

The hydrogen (H$\beta$), magnesium (Mgb) and iron (Fe5270 and Fe5335) lines were taken as indicators of age, metallicity and $\alpha$-enhancement, and were measured using the Lick/IDS index definitions \citep{Worthey_1994} with the \textsc{indexf} software of \citet{Cardiel_2010}. The uncertainties on these measurements were estimated within the software using simulations based on the effect of uncertainties in the line-of-sight velocities and the S/N measured from the spectrum itself. Emission lines were clearly visible in the spectra of the nucleus and the host galaxy of FCC~207 and FCC~306, leading to systematically lower measurements for the H$\beta$ absorption line strength, which would affect the estimates for the age of the stellar populations present. Therefore, to correct for the emission lines, the spectra of both components were modelled again using pPXF to obtain best-fitting models for the absorption and emission spectra, which were used to correct for the emission lines in the decomposed spectra. The line strengths were then converted into estimates of luminosity-weighted age and metallicity by interpolation from the Single Stellar Populations (SSP) model grids of \citet{Vazdekis_2010} using the \textsc{rmodel} software \citep{Cardiel_2003}. These models use the MILES stellar library \citep{Sanchez_2006} to create predictions of the line strengths for single stellar populations covering a wide range in age and metallicity. The MILES webtool\footnote{http://www.iac.es/proyecto/miles/pages/webtools/tune-ssp-models.php} allows the model spectra to be convolved with a Gaussian of the appropriate dispersion to reproduce the spectral resolution of the data, thus producing model predictions that have been matched to the resolution of the data. This step minimizes the loss of information that normally occurs when degrading the data to match lower-resolution models. However, in the sample of galaxies used for this study, only FCC~202 shows any significant rotation that can be measured by MUSE, and so the differences in the models used for each galaxy are negligible.

The H$\beta$ absorption feature was used as an indicator of age, and the metallicity was measured using the combined metallicity index,
\begin{equation} 
	\text{[MgFe]$'$}=\sqrt{\text{Mg}b\ (0.72 \times \text{Fe}5270 + 0.28 \times \text{Fe}5335)},
	\label{eq_MgFe}
\end{equation}
due to its negligible dependence on the $\alpha$-element abundance \citep{Gonzalez_1993, Thomas_2003}. 
An example of the line strengths of the NSC and host galaxy of FCC~222 overplotted onto the SSP model predictions is given in Fig.~\ref{fig:SSP_model}. In this example, the NSC contains younger stellar populations than the host galaxy, indicating that the most intense episode of star formation in the recent past mainly occurs within the NSC region.

To further demonstrate the strength of using \textsc{buddi} to cleanly extract the spectrum of an NSC and obtain estimates of its stellar populations, the line strengths measured from the kinematically obliterated datacube after applying the Voronoi binning technique are also plotted in Fig.~\ref{fig:SSP_model}. These points are the smaller dots, which have been colour-coded according to distance from the centre of the galaxy, which was taken to be the centre of the NSC. The data point for the binned spectrum covering the core of the galaxy where the NSC can be seen is highlighted with a larger circle overplotted. In general, the measurements from the binned spectra agree with those of the host galaxy even into the core of the galaxy, with the bins at lower radii trending towards younger ages and the central bin lying closest to the measurement for the NSC. Consequently, one can see that without modelling the light from the host galaxy, any spectra extracted from the region of the NSC will be dominated by the light from the rest of the galaxy.

\begin{figure}
\begin{center}
\includegraphics[angle=0,width=0.9\linewidth]{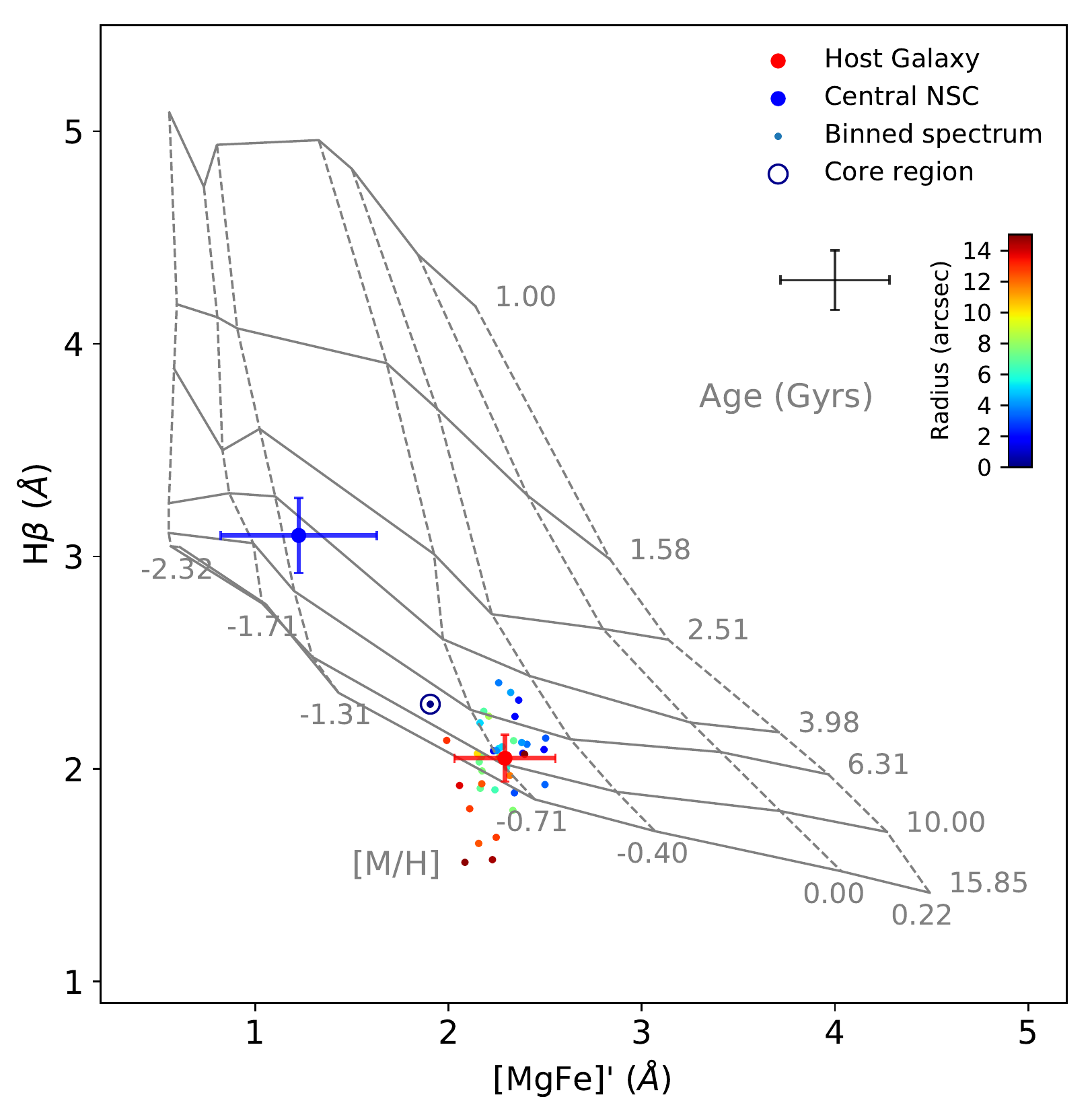}
\caption{An example of the line strength measurements from the nucleus (blue) and host galaxy (red)  of FCC~222 over-plotted onto SSP models to give estimates of the luminosity-weighted ages and metallicities of their stellar populations. For reference, the measurements from the original datacube after smoothing the kinematics and applying the Vornoi binning have been plotted as the small points, colour-coded from blue to red to indicate increasing distance from the centre of the galaxy as indicated by the colour bar. The small data point with the large circle around it represents the binned spectrum covering the core of the galaxy, where the NSC lies, and the error bar below the legend gives the mean error on the measurements from the binned spectra.
 \label{fig:SSP_model}}
\end{center}
\end{figure}

\begin{figure}
\begin{center}
\includegraphics[angle=0,width=0.98\linewidth]{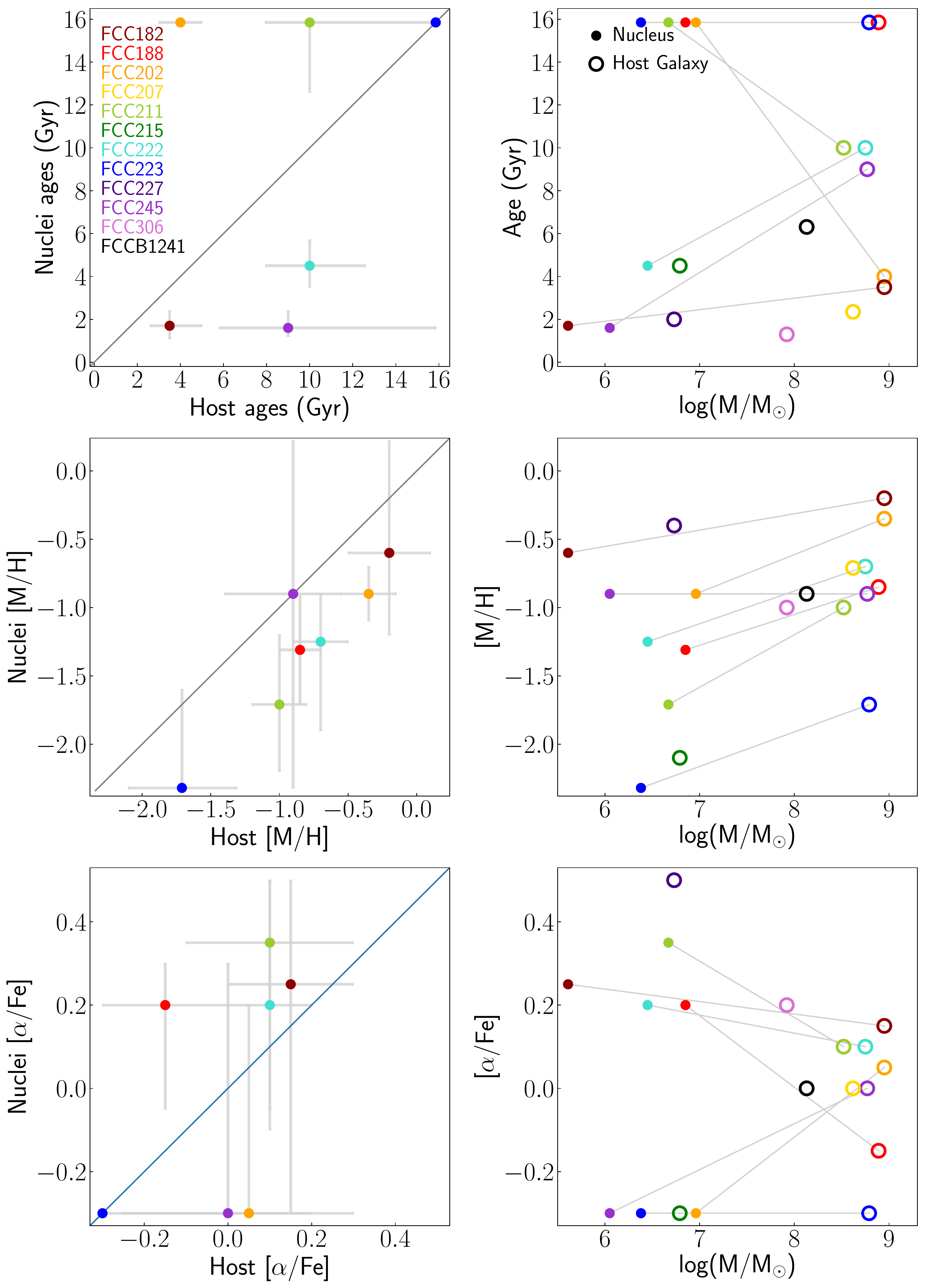}
\caption{A comparison of the luminosity-weighted ages (top), metallicities (middle) and $\alpha$-enhancements (bottom) of the nuclei and their host galaxies, plotted against each other in the left column and against the mass of each component in the right column.  In the left column, the values are only plotted for galaxies where the NSC has sufficient S/N to obtain an estimate of the stellar populations, and the error bars represent the 1-$\sigma$ uncertainties in the estimates. In the right column, the closed and open circles represent the NSCs and the host galaxies respectively, and the grey lines connect the components from the same galaxy. Open circles without a grey line represent host galaxies in which the S/N of the NSC spectrum was too low for this analysis. \label{fig:LW_plots}}
\end{center}
\end{figure}

The luminosity-weighted ages and metallicities of all the nuclei and their corresponding host galaxies are plotted in Fig.~\ref{fig:LW_plots}. Where the measurements or the limits of their uncertainties fell off the model grids, the values were extrapolated to the closest point on the grid. The NSCs of FCC~207, FCC~215, FCC~227, FCC~306 and FCCB~1241 have been omitted from these plots due to too low S/N  ($<10$)} for the stellar populations analysis. In these cases, either the observations were too shallow or the galaxies and their NSCs had lower surface brightnesses than the rest of the sample, leading to lower S/N in their extracted spectra. As a result, only the measurements from the host galaxy spectra in these galaxies have been included in these plots. 

One clear trend that appears in Fig.~\ref{fig:LW_plots} is that while all the NSCs and galaxies display a wide range in metallicity, the NSCs have systematically lower metallicities than their corresponding host galaxies.  This result is in agreement with the trend seen by \citet{Ordenes_2018b} for nuclei within this mass-regime and \citet{Fahrion_2020} for MUSE observations of NSCs in two dwarf galaxies in Centaurus~A (NGC\,5128). Since in-situ star formation fuelled by enriched gas from the rest of the galaxy would produce young stellar populations with similar luminosity-weighted metallicities to the host galaxy, the lower metallicities in the NSCs may indicate that such in-situ star formation only plays a minor role in their later mass-assembly, or that the gas was enriched by earlier star formation within the galaxy.

The ages of the NSCs appear to cover a wide range, likely reflecting the diversity in the times since they formed or last experienced star formation. Three galaxies, FCC~182, FCC~222 and FCC~245, contain younger stellar populations in their NSCs, while the other NSCs whose spectra had sufficiently high S/N were found to be very old. In the cases of FCC~223 and FCC~188, both the NSC and host galaxy were found to be old enough that no distinction between their ages was possible with the data available. As an additional measure on the properties of the host galaxies, the line strengths of the binned spectra from the original datacubes were also measured and plotted onto the SSP model grids. In all cases, these measurements were found to be consistent with those of the host galaxy spectra extracted by \textsc{buddi}. Again, FCC~223 and FCC~188 were found to be uniformly old, and neither galaxy showed evidence of emission, which would act to artificially skew the H$\beta$ line strengths to lower values. Consequently, we conclude that both components within these galaxies are too old to distinguish their ages through line strength fitting.

\subsection{$\alpha$-enhancement}\label{sec:line_strengths2}
Studying the $\alpha$-element enhancement of a stellar population can provide information about the timescale over which those stars were created and the origin of the gas that fuelled the star formation. The $\alpha$-enhancement is measured through the [$\alpha$/Fe] ratio, where $\alpha$-elements and Fe are ejected into the interstellar medium by Type~II  and Ia supernovae (SNeII and SNeI), respectively. Since SNeI start to occur $\sim\!1$\,Gyr after the onset of star formation while SNeII appear much sooner, the ratio of $\alpha$ to Fe-peak elements, such as magnesium to iron, can be used to estimate relative star-formation timescales. For example, a shorter episode of star formation will result in an $\alpha$-enhanced stellar population due to the enrichment of magnesium from the SNeII, and the $\alpha$-enhancement will begin to drop after SNeI appear due to the dilution of magnesium with iron in the interstellar medium \citep[e.g.][]{Matteucci1994}. 

The Mgb and $\langle \text{Fe} \rangle=(\text{Fe}_{5270}+\text{Fe}_{5335})/2$ line strengths from the nuclei and host galaxy spectra were plotted onto the SSP model grids of \citet[][hereafter TMJ models]{Thomas_2011}, who plot the [$\alpha$/Fe] ratio and metallicity for a range of ages of stellar populations. For each nucleus and host galaxy spectrum, the model was selected that corresponded most closely to the age of that structure from the H$\beta$ vs.~[MgFe]$'$ models discussed in Section~\ref{sec:line_strengths1}. 

The [$\alpha$/Fe] ratio for each spectrum was estimated through interpolation from the selected model, and the results are presented in Fig.~\ref{fig:LW_plots}. Due to the low metallicities of the NSCs and their host galaxies, the majority of the line strengths fall within the narrow end of the TMJ models, and the uncertainties in these measurements span a wide range of [$\alpha$/Fe] ratios. As a result, the error bars for these measurements in Fig.~\ref{fig:LW_plots} are very large, and little can be said for certain about the $\alpha$-enhancement of the NSCs and their host galaxies.



\section{Analysis: Mass-weighted stellar populations}\label{sec:analysis3}
The luminosity-weighted stellar populations analysis presented above assumes only a single star-formation even and thus only provides information on the most recent episode of star formation within each component since the stars created during that event tend to be brighter and dominate the light from the galaxy, even when they account for only a small fraction of the mass. As a result, they are best used to determine how long ago the most recent star-formation activity ceased, and provide limited information on the full star-formation history of the galaxy. Consequently, in order to better understand the mass assembly of the NSCs and their host galaxies, and thus distinguish between the proposed formation scenarios, one must look in more detail at the star-formation histories of the NSC and the host galaxies. For this analysis, full spectral fitting was used to derive the mass-weighted stellar populations of each component.

The mass-weighted ages and metallicities were derived from each NSC and host galaxy spectrum using pPXF by applying a regularized fit to the spectrum using a linear combination of template spectra of known relative ages and metallicities. We used the MILES stellar library as template spectra, covering a range in ages of 0.06-18~Gyrs and a range in metallicity of [M/H]~$=-2.32$ to $+0.22$.~The linear regularization process of \citet{Press_1992} was used to smooth the variation in the weights of templates of similar ages and metallicities, where the regularization parameter determined the degree of smoothness required. A multiplicative Legendre polynomial of value 10 was used to correct the shape of the continuum in these fits while making the fit insensitive to dust reddening and without the requirement of a reddening curve \citep{Cappellari_2017}. It should be noted that this smoothing may not completely reflect the true star-formation history of the galaxy, which is likely to be stochastic and variable over short timescales. Instead, it acts to reduce the age-metallicity degeneracy between spectra, thus allowing a more consistent comparison of systematic trends in their star-formation histories.

An unregularised fit was first applied to each spectrum to measure the $\chi^2$ value for that fit. The noise spectrum was then scaled until the $\chi^2/N_{\rm DOF}=1$, where $N_{\rm DOF}$ is the number of degrees of freedom in the fit, or the number of unmasked pixels in the input spectrum. The spectrum was then remodelled using increasing values for the regularization parameter until the $\chi^2$ value had increased by $\Delta \chi^2=\sqrt{2N_{\rm DOF}}$. This criterion identifies the boundary between a smooth fit that still reflects the original spectrum, and a fit that has been smoothed too much and no longer represents the star-formation history of the galaxy.  Since the template spectra are modelled for an initial birth cloud mass of 1 Solar mass,  the final weight of each template reflects the `zero-age' mass-to-light ratio of that stellar population. Therefore, the smoothed variation in the weights of neighbouring template spectra in the final regularised fit represents a simplified star-formation history of that part of the galaxy by defining the relative mass contribution of each stellar population \citep[see e.g.][]{McDermid_2015}. 

\begin{figure}
\begin{center}
\includegraphics[angle=0,width=0.98\linewidth]{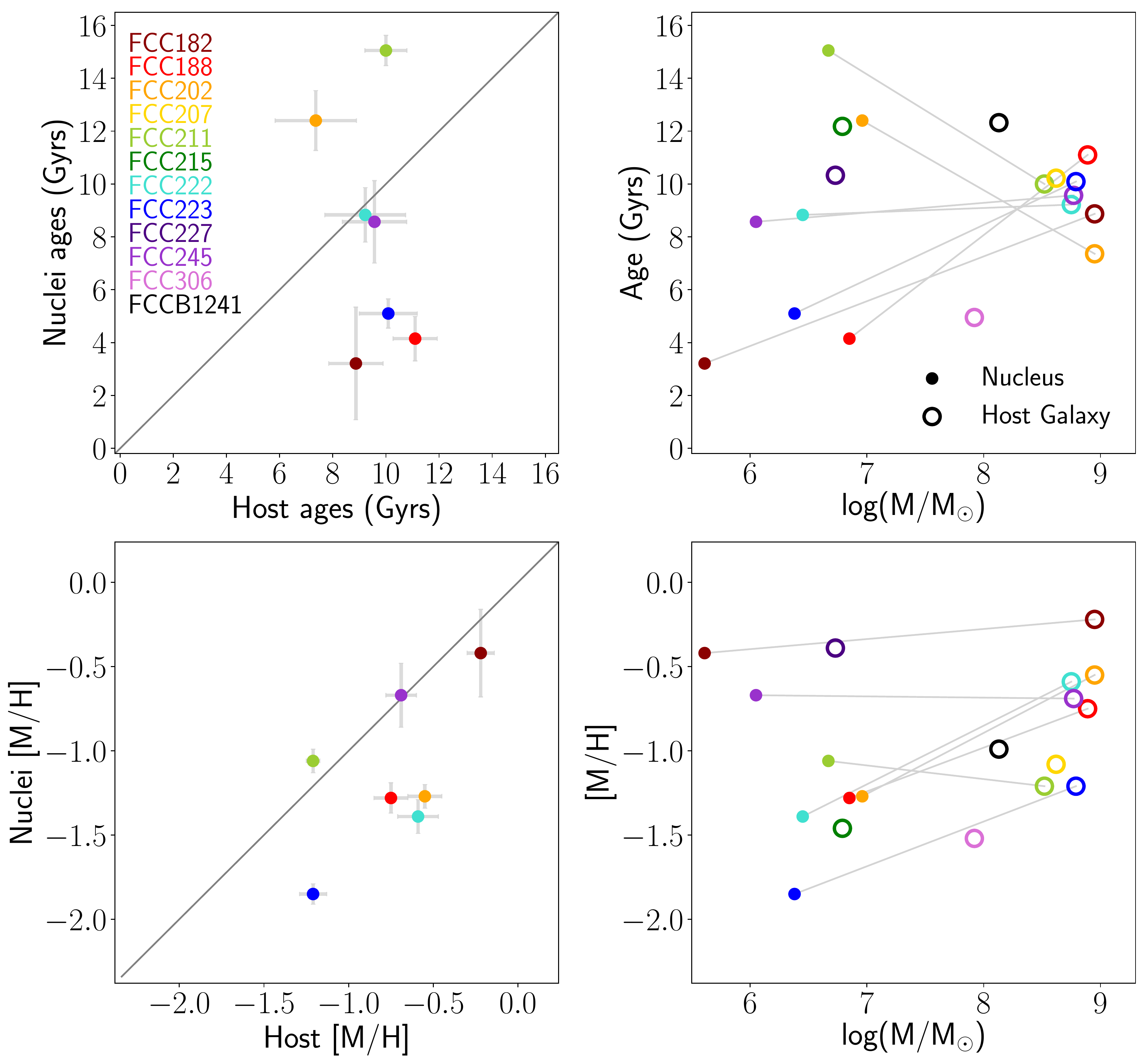}
\caption{ A comparison of the mass-weighted ages (top) and metallicities (bottom) of the nuclei and their host galaxies, plotted against each other in the left column and against the total mass of each component in the right column.  \label{fig:mag_plots2}}
\end{center}
\end{figure}

In many of the spectra, particularly those with higher levels of noise, it was found that the regularized fit was able to find a perfectly smooth weight distribution before $\Delta \chi^2=\sqrt{2N_{\rm DOF}}$. Therefore, in order to be conservative and reduce the chances to applying excessive smoothing, we followed the approach of \citet{Shetty_2015} and reduced the desired $\Delta \chi^2$ to $\Delta \chi^2\!=\!\sqrt{2N_{\rm DOF}}/6$. This value was found to perform well for the low-S/N spectra while not changing the results for high S/N examples significantly when compared to using $\Delta \chi^2=\sqrt{2N_{\rm DOF}}$.

The mean mass-weighted ages and metallicities of each component were then calculated from the weights of each template spectrum using 
\begin{equation} 
	\text{log(Age$_{\text{M-W}}$)}=\frac{\sum \omega_{i} \text{log(Age$_{\text{template},i}$)}}{\sum \omega_{i}}
	\label{eq:age}
\end{equation}
and 
\begin{equation} 
	\text{[M/H]$_{\text{M-W}}$}=\frac{\sum \omega_{i} \text{[M/H]}_{\text{template},i}}{\sum \omega_{i}}
	\label{eq:met}
\end{equation}
respectively, where $\omega_{i}$ represents the weight of the $i^{th}$ template (i.e. the value by which the $i^{th}$ template stellar template is multiplied to best fit the galaxy spectrum), and [M/H]$_{\text{template},i}$ and Age$_{\text{template},i}$ are the metallicity and age of the $i^{th}$ template respectively. A comparison of the mass-weighted ages and metallicities are plotted in Fig.~\ref{fig:mag_plots2}. The uncertainties in these measurements were calculated through a series of simulations, where random noise was added to the best-fitting model to each spectrum to replicate the S/N of the original spectrum for each component. One hundred simulated spectra were created for each component, and the mass-weighted ages and metallicities were measured in the same way with the same values for the regularization. The uncertainties were then calculated as the standard deviation in the measurements from these simulations.

\begin{figure*}
\begin{center}
\includegraphics[angle=0,width=0.8\linewidth]{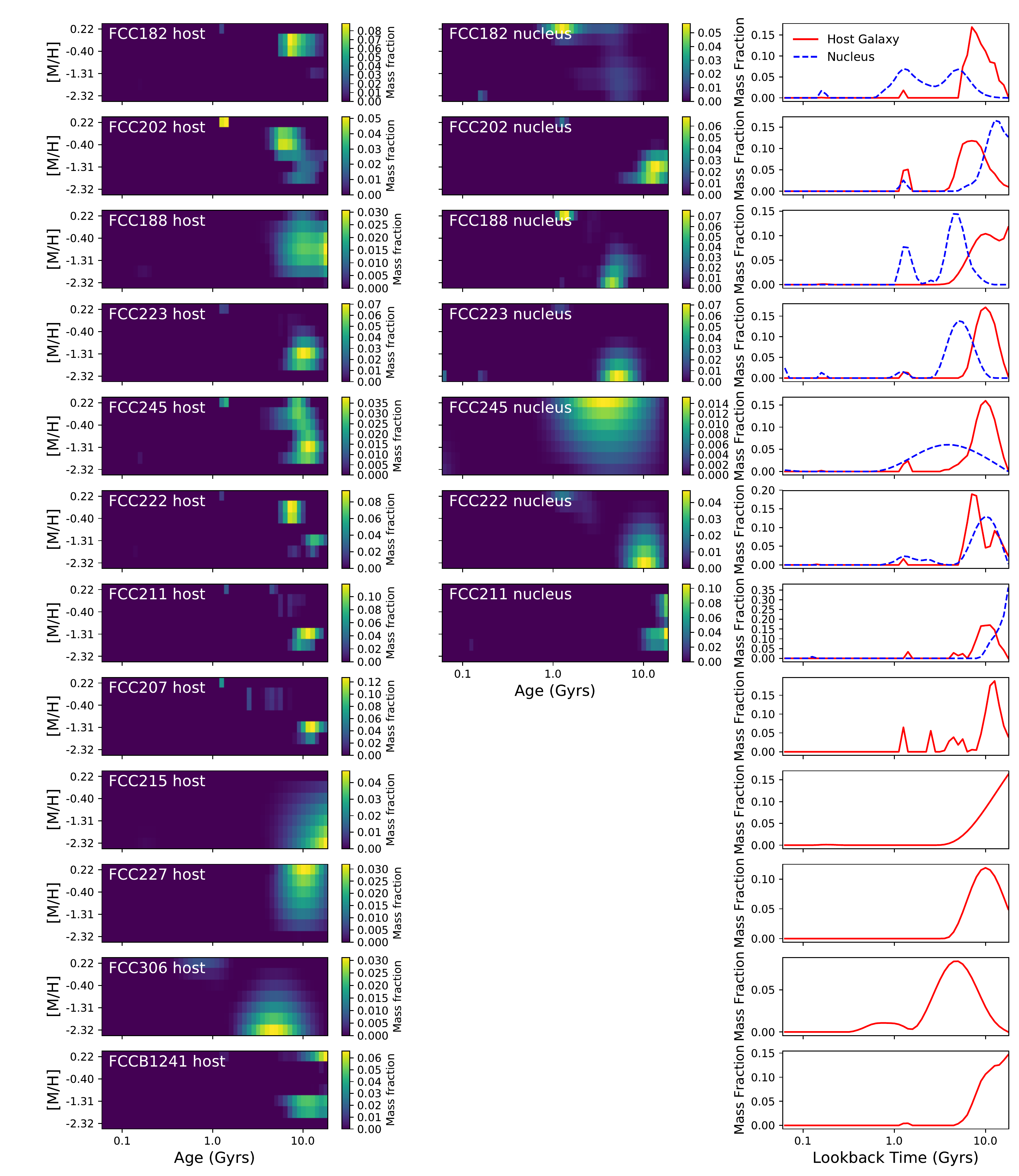}
\caption{The smoothed age-metallicity weights maps for the fits to the nuclei and host galaxy spectra (left and middle columns respectively), with the colourbar next to each plot giving the mass fraction corresponding to each value for age and metallicity. The nuclei for FCC~215, FCC~227 and FCCB~1241 have been omitted due to the low S/N of those spectra leading to poor fits. On the right is the mass fraction created as a function of the lookback time for each component. 
 \label{fig:weights}}
\end{center}
\end{figure*}

It can be seen that the host galaxies are all relatively old, with mass-weighted ages of over 7~billion years. The NSCs, on the other hand, show a range of mass-weighted ages, from $\sim$3 to 13~billion years, and a wide range in metallicities. In general, these results again reflect the scenario that the NSCs and host galaxies were not formed in single star-formation events, but that each component in each galaxy has evolved differently over time.  The metallicities of all the NSCs still appear to be consistent with or to be lower than those of their host galaxies, which may reflect that the majority of their mass was created early in the lifetime of the galaxy before the metallicity built up. The younger stellar populations in some of the NSCs may then reflect that the majority of the mass of the nucleus was  built up later through a recent merger with an infalling younger globular cluster or via star formation fuelled by infalling gas. The old mass-weighted ages of the host galaxies also indicate that the majority of their mass was created long ago, with more recent episodes of star formation triggered by minor mergers, gas accretion from neighbouring galaxies or re-accretion or previously expelled gas contributing insignificantly to the mass of the galaxy.

To further investigate these scenarios, the weights of each template spectrum used in the regularized fits were plotted to create  smoothed age-metallicity weights maps reflecting the mass assembly history of each galaxy. These maps are presented in Fig.~\ref{fig:weights} for each NSC (with sufficient S/N) and host galaxy, with the galaxies ordered by total mass. All but the two faintest galaxies, FCC~215 and FCC~227, show evidence of multiple episodes of star formation, and in all galaxies the majority of the mass was assembled more than 8~Gyrs ago and with more recent star-formation activity contributing small amounts of mass. This result could arise through minor mergers, accretion of material from nearby galaxies or re-accretion of previously expelled gas during the lifetime of these dwarf galaxies, thus building up their mass and pushing them closer to the transition region with ETGs on the mass-size relation. This theory is consistent with the hierarchical formation scenario, where the mass of dwarf galaxies is built up through multiple minor mergers or accretion of matter, and would also explain the single episode of star formation that created the two lowest-mass galaxies if they are simply less evolved.

The NSCs also all appear to show evidence for multiple episodes of star formation during their lifetimes. The majority of the NSCs host an old and metal-poor component with secondary peaks in their star formation resulting in younger, more metal-rich stellar populations. This result is consistent with the majority of the mass of the NSC forming long ago from inwardly-migrating GCs and mergers with other infalling GCs, with more recent in-situ star formation within the NSC being fueled by enriched gas. This scenario would also explain the blue cores seen in some dwarf galaxies \citep[e.g.][]{Ulrich_2017,Chung_2019}, and the ongoing star formation detected in the nuclear region of FCC~207. Even the star formation histories of FCC~245 and FCC~211, which show more extended star formation in in Fig.~\ref{fig:weights}, may also reflect multiple short episodes of star formation that have been smoothed out in the regularized fits.

%

\section{Summary and Discussion}\label{sec:discussion}
There are two main formation scenarios for dwarf galaxy NSCs. In the in-situ scenario, gas distributed throughout the galaxy is funnelled into the central regions, where it eventually triggers an episode of star-formation activity and forms a central nuclear star cluster. The migration scenario, on the other hand, suggests that the NSC originates as a globular cluster  in the outskirts of the galaxy and then migrates into the central regions through dynamical friction. Mergers with other infalling globular clusters could contribute to the growth of the nucleus over time through dry mergers. 

While the formation scenario is often considered to be linked to the mass of the host galaxy, with migration of GCs dominating in low-mass galaxies while in-situ star formation becomes more common at higher masses, it is becoming increasingly clear that no single process is completely responsible for the formation and mass assembly of NSCs. In response to these findings, a third hybrid scenario has been proposed- the wet-migration scenario. In this case, a gas-rich massive star cluster forms within the galaxy and continues forming stars as it migrates into the centre of the galaxy to create the nucleus. As with the migration scenario, this scenario also proposes that later wet mergers with infalling gas-rich massive clusters would contribute to building up the mass of the nucleus.  Consequently, it is important to no longer only determine the dominant processes that built up the mass of the NSC, but instead to consider the relative importance of the dissipational and dissipationless processes as a function of mass and environment.

Each of these three scenarios would leave different signatures in the stellar populations of the NSC. For example, evidence of multiple episodes of star formation could indicate that the NSC has built up its mass through mergers with other infalling gas-rich \mbox{YMSCs}, whereas mergers with old GCs would result in less diversity in the stellar populations as they would be too similar to be disentangled. Additionally, higher metallicities in the NSCs than the host galaxies may reflect that the NSC formed from infalling gas after the star formation in the rest of the galaxy was truncated. Therefore, by studying the star formation histories of both components in these galaxies we can better understand how they have formed and evolved over time. In this work, we studied the stellar populations of the NSCs and their host galaxies in a sample of dwarf galaxies in the Fornax cluster observed with the MUSE IFU spectrograph to determine which of these scenarios contributed towards their formation. In order to reduce contamination due to the superposition of light from the host galaxies upon these faint structures, we  used a new technique to model the NSC and the host galaxy spheroid light component as a function of wavelength within the MUSE datacube, cleanly decoupling their spectra and allowing independent analysis of their star-formation histories.

The aim of this study was to use \textsc{buddi} to model the NSC and the host galaxy in a sample of dwarf galaxies in Fornax with observations using the MUSE IFU spectrograph in order to cleanly extract their spectra and investigate their stellar population properties. The first step in the analysis was to calculate the redshifts of the NSCs and thus spectroscopically confirm that they lie at the kinematic core of their host galaxies and thus are true NSCs \citep{Neumayer_2011}. Of the 12 galaxies, 11 of the NSCs were found to be consistent with the kinematic core of their host, with the exception of FCC~306 where the NSC velocity was $\sim\!3$\,$\sigma$ apart from that of the host galaxy. In this case, the proposed NSC is, therefore, likely to be a star-forming region or GC within the galaxy and not the NSC. Furthermore, FCC~222 showed a second bright point source close to its core, which may represent a GC falling into the core of the galaxy to build up the mass of the NSC. Again, the velocity of this object was found to be inconsistent with that of the NSC but still within the velocity dispersion of the host galaxy, thus indicating that it is more likely to be a GC within the galaxy that happens to fall along almost the same line of sight as the NSC.

Having spectroscopically confirmed that the central point sources within each galaxy were NSCs, the galaxies were modelled with \textsc{buddi} to create wavelength-dependent models of each light-profile component from which their spectra could be cleanly extracted. The luminosity and mass-weighted stellar populations of the NSCs and their host galaxies were estimated from these spectra using both line strength analysis and full spectral fitting respectively, where the central PSF component was assumed to reflect the NSC and all other components were combined to represent the host galaxy. The luminosity-weighted stellar populations show a wide distribution in ages of both the NSCs and their host galaxies, with no clear trend between the ages of components within the same system. The metallicities of the NSCs, however, were generally found to be similar or lower than their host galaxies, and are consistent with the photometric findings of \citet{Ordenes_2018b} that NSCs in Fornax Cluster dwarfs of similar masses are generally metal-poor and with a wide distribution in their ages. No single spectroscopic study exists in the literature giving the metallicites of NSCs and their host galaxies over a wide range of masses, but \citet{Neumayer_2020} recently collected the spectroscopic metallicities for these structures in galaxies with masses $10^{8}-10^{10}M_{\odot}$ from \citet{Koleva_2009}, \citet{Paudel_2011} and \citet{Kacharov_2018}. They found that in galaxies with masses $<10^{9}M_{\odot}$ the majority of the NSCs have lower metallicities than their host galaxies, while the opposite is true for higher mass galaxies. On the other hand,  \citet{Spengler_2017} found that metallicities of NSCs and their host galaxies in the Virgo Cluster are indistinguishable, and a spectroscopic study of the NSCs in Centaurus~A (NGC\,5128) dwarf galaxies by \citet{Fahrion_2020} found two cases where the NSCs have lower metallicities than their host galaxies.

Since the light from a galaxy can be dominated by  even a small mass fraction of young, recently formed stars, the luminosity-weighted stellar populations only  provide information on the more recent star formation within the galaxy. In order  to obtain overall star-formation histories of the NSCs and their hosts, the mass-weighted stellar populations were analysed. This analysis revealed that the mean ages of the host galaxies lie within a narrow range, $\sim\!8\!-\!12$\,Gyrs while the NSCs show a much broader distribution. Again, the metallicities of the NSCs were found to be consistent with or lower than their host galaxies, indicating that the trend seen in the luminosity-weighted metallicities is not an effect of recent star formation. Furthermore, the star-formation histories of the host galaxies all show evidence of multiple periods or extended episodes of star formation, which agrees with the lower $\alpha$-enhancements measured from their spectra. Similarly,  many of the nuclei show multiple stellar populations, with only FCC~245 showing a single extended episode of star formation. While care must be taken in interpreting the star-formation histories measured in this way as they only reflect an estimate of the smoothed variation in star formation over the lifetime of the galaxy, it is sufficient to rule out NSCs consisting of a single stellar population and thus indicates that these NSCs have evolved with time.

Multiple stellar populations within NSCs have been found in many previous studies  of dwarf galaxies \citep[e.g.][]{Walcher_2006, Rossa_2006, Seth_2006, Seth_2010, Koleva_2011, Lyubenova_2013}, and likely reflect the mass assembly of the NSC since it formed. For example, \citet{AlfaroCuello_2019} proposed that old, metal-poor stellar populations within NSCs have been created through inward migration of GCs while gas infall is responsible for contributing towards the younger, more metal-rich stellar populations. Furthermore, \citet{Turner_2012}, \citet{den_brok_2014}, \citet{Lyubenova_2019} and \citet{Fahrion_2020} have found that in low-mass galaxies, the dominant process contributing towards the mass assembly of the NSC is mergers with infalling GCs, while gas inflow and the subsequent star formation are a secondary but still significant process. One would expect that any new stars formed through in-situ star formation fuelled by infalling gas from the rest of the galaxy or through mergers with YMSCs containing  gas and ongoing star formation would have similar metallicities to the host galaxy, and thus that these young, bright stars would dominate the luminosity-weighted stellar populations. GCs on the other hand generally have lower metallicities than their host galaxies, representing the metallicity of the host galaxy when the GC was formed \citep{Jordan_2004, Puzia_2005}. As a result, mergers with migrating GCs would result in a more massive NSC with a lower metallicity than the host galaxy. Consequently, the results presented in this work point to a scenario where the NSCs in Fornax dwarf galaxies originated as GCs that migrated into the core of the galaxy and which have built up their mass mainly through mergers with infalling GCs, with in-situ star formation fuelled by inflowing gas or wet mergers with YMSCs playing a more minor role.

This work presents a pilot study of a sample of nucleated dwarf galaxies within the Fornax Cluster that were observed with MUSE, and demonstrates the strength of using \textsc{buddi} to separate the spectra of each component in order to better understand their star formation histories with minimal contamination. There remains plenty of scope for future work following on from this project, such as studying the detailed chemical composition of NSCs, in particular their [$\alpha$/Fe], [C/Fe], [C/N] ratios and comparing the structural parameters and stellar populations of each component within nucleated and non-nucleated dwarfs to better understand the processes that initially form the NSC, and extracting the spectra of GCs within each galaxy to better understand how they relate to the NSCs.~Furthermore, UCDs have been proposed to be the stripped remnants of nucleated dwarf galaxies, and this theory can be studied in more detail with this new technique to isolate the properties of the NSCs. While these ideas are beyond the scope of this work, together they will contribute to a better understanding of how NSCs form and evolve, and their role in galaxy evolution.


\section*{Acknowledgements}
\addcontentsline{toc}{section}{Acknowledgements}
We would like to thank the anonymous referee for their favourable report. The authors would like to thank Shravan Shetty for a useful discussion on the regularized fits with pPXF.
This study was based on data obtained from the ESO Science Archive Facility under the following request numbers: P094.B-0576 (PI: Emsellem), P094.B-0895 (PI: Lisker),  P096.B-0063 (PI: Emsellem), P096.B-0399  (PI: Napolitano), P097.B-0761 (PI: Emsellem), P098.B-0239 (PI: Emsellem), P101.C-0329 (PI: Vogt) and P102.B-0455 (PI: Johnston).
This study also used data obtained with the Dark Energy Camera (DECam), which  was constructed by the Dark Energy Survey (DES) collaboration.
This project is supported by FONDECYT Regular Project No.\,1161817 and the BASAL centre for Astrophysics and Associated Technologies (PFB-06). 
The authors acknowledge support from CONICYT project Basal AFB-170002. 
E.J.J.~acknowledges support from FONDECYT Postdoctoral Fellowship Project No.\,3180557. 
T.H.P.~acknowledges ECOS-Sud/CONICYT project C15U02.
P.E.~acknowledges support from the Chinese Academy of Sciences (CAS) through CAS-CONICYT Postdoctoral Fellowship CAS150023 administered by the CAS South America Center for Astronomy (CASSACA) in Santiago, Chile.
Y.O.-B.~acknowledges financial support through CONICYT-Chile (grant CONICYT-PCHA/Doctorado Nacional No.~2014-21140651) and the DAAD through the PUC-HD Graduate Exchange Fellowship.
Y.R.~acknowledges support from FONDECYT Postdoctoral Fellowship Project No.~3190354 and NSFC grant No.\,11703037. 
S.M. thanks PSB for Hotspot.
M.A.T.~acknowledges support by the Gemini Observatory, which is operated by the Association of Universities for Research in Astronomy, Inc., on behalf of the international Gemini partnership of Argentina, Brazil, Canada, Chile, the Republic of Korea, and the United States of America.


\appendix

\section{Analysis: simulations of dwarf galaxy datacubes }\label{sec:tests}

In order to test the reliability of the results derived from the spectra extracted by \textsc{buddi} for the NSCs, a final test was carried out by creating a series of model datacubes with similar properties to the data used in this study and extracting the spectra in the same way.

As a starting point, a model image of a galaxy was created, using the fit parameters derived through the fit to the white light image of FCC~245 to create a model 2-component galaxy with a S\'ersic profile representing the host galaxy and a PSF profile for the NSC. The integrated magnitudes for these two structures were changed however, using the $i$-band total magnitudes from FCC~211 and FCC~215 for the host galaxy and NSC respectively as these components lie close to the median values for these components in the corresponding distributions in Table~\ref{tab:masses}. 

This model image was then converted into a series of datacubes by replacing each pixel with a spectrum representing the integrated light from each component in the proportions determined from the model image. The spectrum for each component was taken as a single stellar population from the MILES template spectra, and the 10 model datacubes include cases where the NSC contains younger, older, more metal-rich, more metal-poor and similar stellar populations relative to the host galaxy.

\begin{figure}
\begin{center}
\includegraphics[angle=0,width=0.9\linewidth]{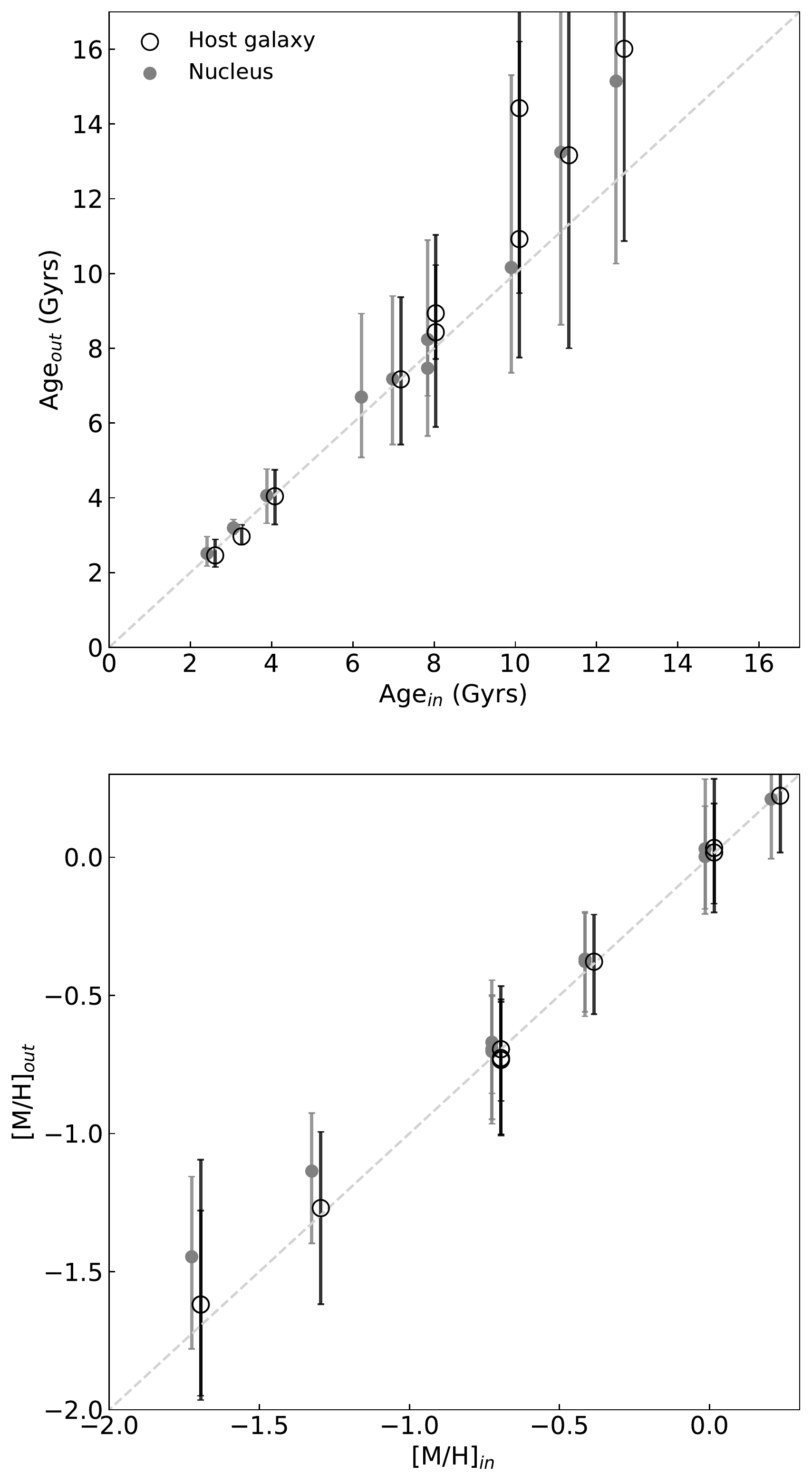}
\caption{Comparison of the luminosity-weighted ages (top) and metallicities (bottom) derived from the simulated datacubes for the host galaxies (hollow black circles) and the NSCs (filled grey circles) against the stellar populations used to create those mock galaxies. The input measurements have been offset to slightly negative and positive values in order to better display the results from the NSCs and the host galaxies respectively. 
 \label{fig:test_cubes}}
\end{center}
\end{figure}

Having created the model datacubes, a sensible level of noise was required to achieve a realistic test. To best simulate the noise and potential contamination from foreground and background light sources, the residual sky and residual datacubes for FCC\,245 were combined with each model galaxy datacube. These cubes were selected due to the presence of a bright star within the FOV which dominates the integrated light from the datacube, and thus was an additional test for contamination from the light that was not masked out. Finally, these galaxies were each run through \textsc{buddi} in the same way as the real MUSE data, and the luminosity-weighted stellar populations measured. The results are presented in Fig.~\ref{fig:test_cubes}, showing a comparison between the stellar populations that were used to create the simulated datacubes and the line strength measurements from the extracted spectra for each component. It can be seen that the results are consistent, indicating that the stellar populations derived from the spectra extracted by \textsc{buddi} for the host galaxy and NSC are reliable within the accuracy of the model spectra. The larger offsets between the input and output values in some of the models are due to the small differences in line strengths in the old, metal-poor region of the SSP models.

\bsp	
\label{lastpage}
\end{document}